\documentclass[aps,floatfix,showpacs,superscriptaddress,showkeys]{revtex4}
\usepackage{amssymb,amsmath,amsfonts,latexsym,graphicx,epsfig,here}
\usepackage{tabularx}
\usepackage{titlesec}
\usepackage{color}
\usepackage{float}
\titleformat{\section}{\large\bfseries}{\thesection}{1em}{}

\newcommand{\bea}{\begin{eqnarray}}
\newcommand{\ena}{\end{eqnarray}}
\newcommand{\be}{\begin{equation}}
\newcommand{\en}{\end{equation}}
\newcommand{\nn}{\nonumber\\}

\newcommand{\la}{\langle}
\newcommand{\ra}{\rangle}

\newcommand{\tH}[1]{|\widehat{H}_{#1}|^2}

\begin{document}

\hfill MITP/18-039 (Mainz)

\title{Nonleptonic two-body decays of single heavy baryons 
$\Lambda_Q$, $\Xi_Q$, and $\Omega_Q$ $(Q=b,c)$ \\
induced by $W$ emission in the covariant confined quark model}  

\author{Thomas Gutsche}
\affiliation{Institut f\"ur Theoretische Physik, Universit\"at T\"ubingen,
Kepler Center for Astro and Particle Physics, 
Auf der Morgenstelle 14, D-72076 T\"ubingen, Germany}
\author{Mikhail~A.~Ivanov}
\affiliation{Bogoliubov Laboratory of Theoretical Physics,
Joint Institute for Nuclear Research, 141980 Dubna, Russia}
\author{J\"urgen~G.~K\"orner}
\affiliation{PRISMA Cluster of Excellence, Institut f\"{u}r Physik,
Johannes Gutenberg-Universit\"{a}t, D-55099 Mainz, Germany}
\author{Valery E. Lyubovitskij}
\affiliation{Institut f\"ur Theoretische Physik, Universit\"at T\"ubingen,
Kepler Center for Astro and Particle Physics,
Auf der Morgenstelle 14, D-72076 T\"ubingen, Germany}
\affiliation{Departamento de F\'\i sica y Centro Cient\'\i fico
Tecnol\'ogico de Valpara\'\i so-CCTVal, Universidad T\'ecnica
Federico Santa Mar\'\i a, Casilla 110-V, Valpara\'\i so, Chile}
\affiliation{Department of Physics, Tomsk State University,
634050 Tomsk, Russia}
\affiliation{Laboratory of Particle Physics, Tomsk Polytechnic University,
634050 Tomsk, Russia}

\begin{abstract}

We have made a survey of heavy-to-heavy and heavy-to-light nonleptonic
heavy baryon two-body decays and have identified those decays 
that proceed solely via $W$-boson emission, i.e. via the  
tree graph contribution. Some sample decays are 
$\Omega_{b}^{-}\to\Omega_{c}^{(*)0}\rho^{-}(\pi^{-}),\,
\Omega_{b}^{-}\to\Omega^{-}J/\psi(\eta_{c}),\,
\Xi_{b}^{0,-}\to\Xi^{0,-}J/\psi(\eta_{c}),\, 
\Lambda_{b}\to \Lambda J/\psi(\eta_{c}),\,
\Lambda_{b}\to \Lambda_{c} D_{s}^{(\ast)},\,
\Omega_{c}^{0}\to\Omega^{-}\rho^{+}(\pi^{+})$ 
and $\Lambda_c \to  p \phi$. 
We make use of the covariant confined quark model previously developed by us 
to calculate the tree graph contributions to these decays. 
We calculate rates, branching fractions and, for some of these decays, 
decay asymmetry parameters. We compare our results to experimental findings
and the results of other theoretical approaches when they are available.
Our main focus is on decays to final states with a lepton pair because of 
their clean experimental signature.  
For these decays we discuss two-fold polar angle decay distributions such 
as in the cascade decay 
$\Omega_{b}^{-}\to\Omega^{-}(\to \Xi\pi,\Lambda K^{-})+J/\psi(\to \ell^{+}\ell^{-})$. 
Lepton mass effects are always included in our analysis.

\end{abstract}

\today



\maketitle

\section{Introduction }

In the present paper we focus on the study of nonleptonic $W$-emission
(also referred to as tree graph contributions~\cite{Leibovich:2003tw})
two-body decays of single heavy baryons 
\be
B_1(q_1\,q_2\,q_3) \to B_2(q'_1\,q'_2\,q'_3) + M(q_m \bar q_{\bar m})
\en
where $M$ stands for a pseudoscalar meson $P$ or a vector meson $V$. One of the
quarks in the initial baryon is heavy.
A necessary condition for the contribution of the tree graph class 
of decays is that a light quark pair $q_{i}q_{j}=q'_{i}q'_{j}$
is shared by the parent and daughter baryon $B_{1}$ and $B_{2}$, respectively.
A sufficient condition for the tree graph class of decays is that
i) $ q_m$ is not among $q_1,\,q_2,\,q_3$ and
ii) $q_{\bar m}$ is not among $q'_1,\,q'_2,\,q'_3$ when $q_m,q_{\bar m}$ are
not among the quarks participating in the weak interaction. 
For completeness we also analyze those nonleptonic modes which 
are contributed to by both tree graph contributions 
and by a specific class of $W$-exchange contribution which vanish in the
$SU(3)$ limit. 

Our investigation is limited to the decays of the lowest lying ground states.
The higher lying ground-state bottom and charm baryons such as the
spin 3/2 partners decay strongly or 
electromagnetically. For these the weak decays do occur but have tiny
branching ratios
and are therefore not so interesting from an experimental point of view.
In this work we therefore concentrate on the weak nonleptonic decays of the
lowest lying spin 1/2
ground state bottom baryons $\Lambda^0_b,\, \Xi^0_b,\,\Xi^-_b,\,\Omega^-_b$
and $\Lambda^+_c,\, \Xi^+_c,\,\Xi^0_c,\,\Omega^0_c$.
 
Based on the necessary and sufficient conditions formulated above
we have made a search for heavy-to-heavy and heavy-to-light nonleptonic
heavy baryon decays that are solely contributed to by the tree graph
contributions. 
We have identified a number of decays in this class.
Some of these decays have been seen or are expected to be seen in the near
future, or upper bounds have been established for them. 
Although some of these tree graph decays are singly or doubly 
Cabibbo–Kobayashi–Maskawa (CKM) suppressed 
the huge data sample available at the LHC on heavy baryon decays 
will hopefully lead to the detection of the remaining decay modes in the near 
future. 

Nonleptonic baryon and meson decays are interesting from the point of view
that they provide a testing ground for our understanding of QCD. They involve
an interplay of short-distance effects from the operator product expansion 
of the 
current-current Hamiltonian and long-distance effects from the evaluation of 
current-induced transition matrix elements. On the phenomenological
side one can investigate the role of $1/N_{c}$ color loop 
suppression effects. The analysis of nonleptonic two-body heavy baryon decays
can also contribute to the determination of CKM matrix elements if the
theoretical input can be brought under control.

An analysis of the tree graph decays provides important information on the 
value of the heavy-to-heavy and heavy-to-light form factors 
$\la B_{2}|J^{\mu}|B_{1} \ra$ at one particular $q^{2}$-value corresponding 
to the mass of the final state meson. A small final-state meson mass implies
that the form factors are probed close to the maximum recoil end of the
$q^{2}$-spectrum where one has reliable predictions
from light cone sum rules. The decay constants describing the 
mesonic part of the transition $\la M|J^{\mu}|0\ra$ are either
experimentally available from weak or
electromagnetic decays of the mesons, or are constrained from further 
theoretical model analysis, lattice calculations or QCD sum rule
analysis.  

In this paper we present a detailed analysis of the nonleptonic
tree graph heavy baryon
decays in the framework of our covariant confined quark model (CCQM)
proposed and developed 
in Refs.~\cite{Ivanov:1996pz}-\cite{Branz:2010pq}. The CCQM model 
is quite flexible in as much as it allows one to study mesons, baryons  
and even exotic states such as tetraquark states as bound states of their 
constituent quarks. In this approach particle 
transitions are calculated through Feynman diagrams involving quark
loops. In the present application the current-induced baryon transitions 
$B_{1}\to B_{2}$ are described by a two-loop diagram which requires a genuine 
two-loop calculation. The vacuum-to-meson transition involves a one-loop 
calculation. The high 
energy behavior of quark loops is tempered by nonlocal Gaussian-type vertex
functions with a Gaussian-type fall-off behavior characterized by a size
parameter specific to the meson and baryon in question. The nonlocal 
hadron-quark vertices have an interpolating current structure. 
In a recent refinement of our model we have incorporated quark
confinement in an effective way~\cite{Branz:2009cd}-\cite{Gutsche:2013pp}. 
Quark confinement is introduced at the level 
of Feynman diagrams and is based on an infrared regularization of the relevant 
quark-loop diagrams. In this way quark thresholds corresponding 
to free quark poles in the Green functions are removed (see details in 
Refs.~\cite{Branz:2009cd}-\cite{Gutsche:2013pp}). 
Tree graph contributions to some of the decays treated in this paper have
been discussed before
in~\cite{Hsiao:2015cda,Hsiao:2017tif,Fayyazuddin:2017sxq}. 

The paper is structured as follows. 
In Sec.~II we provide a list of nonleptonic heavy baryon decays that proceed
solely by the tree graph contribution. We classify the decays according to
their CKM and color structure.
In Sec.~III we discuss the parameters of the effective weak Lagrangians that
induce the current-induced transitions used in this paper.
In Sec.~IV we define the matrix elements for the nonleptonic decays
$1/2^{+}\to 1/2^{+}\,0^-(1^{-})$ and $1/2^{+}\to 3/2^{+}\,0^{-}(1^{-})$ and
relate the helicity amplitudes to the invariant amplitudes of these decay
processes. In this way we obtain very
compact expressions for the total rates.
In Sec.~V we discuss the structure of the interpolating currents that
describe the interactions 
of light and heavy baryons with their constituent quarks.
In Sec.~VI we briefly describe the main features of the CCQM model and present
the values of the model parameters such as the constituent quark masses
that have been determined by a global fit to a multitude of decay processes.
In Sec.~VII we present our numerical results on the partial branching
fractions of the various nonleptonic baryon decays and compare them to
experimental results when they are available. We also discuss angular decay
distributions for the most interesting cascade decay processes and present
numerical results on the spin density matrix elements that describe the
angular decay distributions.
Finally, in Sec.~VIII, we summarize our results.  

\section{Classification of factorizing decays}

In this section we list all nonleptonic two-body decays of the lowest
lying spin-1/2 ground state
heavy bottom and charm baryon decays $\Lambda_{b,c},\Xi_{b,c},\Omega_{b,c}$
which proceed solely via the tree graph. We exclude the higher-lying
ground-state baryons which decay
either strongly or electromagnetically.

The quark content of the baryon and meson states participating in the
nonleptonic two-body decays is given by
\be
B_1(q_1\,q_2\,q_3) \to B_2(q'_1\,q'_2\,q'_3) + M(q_m \bar q_{\bar m}).
\en
One can formulate necessary and sufficient conditions for the
decays which solely proceed via the tree graph contribution. These read
\bea
    {\rm necessary \,condition}: \qquad && q_{i}q_{j}=q'_{i}q'_{j} \quad i,j=1,2,3\nn
    {\rm sufficient\, condition}:\qquad && q_{ m} \ni q_1,\,q_2,\,q_3\,;\quad
    q_{\bar m} \ni q'_1,\,q'_2,\,q'_3 ;\nn
    \qquad \qquad&& q_{ m},\, q_{\bar m} \,{\rm not\, part\,
      of\,the \, effective
      \,current-current\, Lagrangian}
    \ena
    We have made a survey of two-body nonleptonic bottom and charm baryon decays
    which are solely contributed to by the tree diagram. The results of this
    search are given as follows.

\vspace{0.2cm}
{\bf Bottom baryon decays:}
\be
\begin{array}{rcclr}
   &b\to c \quad c \to s:         & {\rm color\ favored}
  &\Lambda_b^0 \to \Lambda_c^+\,D_s^{(*)-}  &
\\
&& {\rm color\ suppressed}    &  \Lambda_b^0 \to \Lambda^0\,\eta_c(J/\psi) &
\\
&& &\Xi_b^{0,-}\to \Xi^{0,-}\,\eta_c(J/\psi) &
\\
&&&\Omega_b^{-}\to \Omega^{-}\,\eta_c(J/\psi) &
\\
&b\to c \quad u \to d:         & {\rm color\ favored}&
\Omega_b^{-}\to \Omega_c^{(*)0}\,\pi^-(\rho^-) &
\\
&&{\rm color\ suppressed}
&\Xi_b^- \to \Sigma^{-}\,D^{(*)0} &
\\
&&&\Omega_b^-\to \Xi^{(*)-}\,D^{(*)0} &
\\
X:\qquad&b\to c \quad c\to d & {\rm color\ favored}&
\Xi_b^0 \to \Xi_c^{+}\,D^{(*)-} &
\\
&&&\Omega_b^-\to \Omega_c^{(*)0}\,D^{(*)-} &
\\
&&{\rm color\ suppressed}&\Lambda_b^0 \to n\,\eta_c(J/\psi) \\
&&&\Xi_b^- \to \Sigma^{-}\,\eta_c(J/\psi) &
\\
&&&\Xi_b^0 \to \Lambda^{0}(\Sigma^0)\,\eta_c(J/\psi) &
\\
&&&\Omega_b^-\to \Xi^{(*)-}\,\eta_c(J/\psi) &
\\
X:\qquad&b\to c \quad u\to s & {\rm color\ suppressed} &\Xi_b^- \to \Xi^{-}\,D^{(*)0}
\\&&  &\Omega_b^- \to \Omega^{-}\,D^{(*)0} &
\\
X:\qquad&b\to u \quad c\to s & {\rm color\ favored } &\Lambda_b^0 \to p\,D_s^{(*)-}  
\\
&&&\Omega_b^-  \to \Xi^{\ast\, 0}\,\,D_s^{(*)-} & {(*)}
\\
& & {\rm color\ suppressed} 
& \Omega_b^- \to \Omega^- \bar D^{(\ast)0} 
\\
X:\qquad&b\to u \quad u\to d & {\rm color\ suppressed } 
&\Xi_b^-  \to \Sigma^-\,\pi^0(\eta,\eta',\rho^0,\omega)  &  
\\
&&&\Omega_b^-\to \Xi^{(*)-}\,\pi^0(\eta,\eta',\rho^0,\omega)  &    
\\
XX:\qquad&b\to u \quad c\to d & {\rm color\ favored } &\Xi_b^0 \to \Sigma^+\,D^{(*)-} &
\\
&&&\Xi_b^- \to \Sigma^0\,D^{(*)-} & {(*)}
\\
&&&\Omega_b^-\to \Xi^{(*)0}\,D^{(*)-} & 
\\
&&&\Omega_b^-\to \Xi^{(*)-}\,\bar D^{(*)0} &
\\
XX:\qquad&b\to u \quad u\to s & {\rm color\ suppressed }
&\Xi_b^- \to \Xi^-\,\pi^0(\eta,\eta',\rho^0,\omega)  &  
\\
&&&\Omega_b^-\to \Omega^{-}\,\pi^0(\eta,\eta',\rho^0,\omega)  &  
\end{array}
\label{bottom}
\en 

{\bf Charm baryon decays:}
\be
\begin{array}{rcclr}
  &c\to s \quad d \to u:         & {\rm color\ favored}
  &\Xi_c^+ \to \Xi^{*\,0}\,\pi^+ (\rho^+) & \hspace*{1.6cm}(*)\\
  &&& \Omega_c^0 \to \Omega^-\,\pi^+(\rho^+) &
  \\
  && {\rm color\ suppressed}
  & \Omega_c^0 \to \Xi^{*\,0}\,\bar K^{(*)0} & \hspace*{1.6cm}(*)
  \\
  X:\qquad &c\to s \quad s \to u: &{\rm color\ suppressed} & \Lambda_c^+ \to p \phi &
  \\
\end{array}
\label{charm}
\en 
We denote singly Cabibbo suppressed decays by one cross ($X$) and
doubly Cabibbo suppressed decays by two crosses ($XX$). There are a number
of decays which are also admitted by a $W$-exchange contribution which are,
however, dynamically suppressed due to the K\"orner-Pati-Woo
theorem~\cite{Korner:1970xq,Pati:1970fg} in the $SU(3)$ limit. These decays
are marked by an
asterix ($*$). We also do not list $W$--emission contributions where the flavor
symmetries of the initial and final spectator quarks do not match, i.e.
where one has $[q_iq_j] \to \{q_i'q_j'\}$ such as in
$\Lambda_b^0 \to \Sigma^0 J/\psi$ or in $\Xi_b^0 \to \Xi_c'^{(*)+}\,D^{(*)-}$. 

\section{Effective Nonleptonic Weak Lagrangians}

In this section we list the effective nonleptonic weak Lagrangians, 
which will be used for the calculation of heavy baryon decays in present paper. 
Here and in the following $V_{q_1q_2}$ 
are Cabibbo-Kabayashi-Maskawa (CKM) matrix elements. 
In our manuscript we use the following set of CKM matrix elements: 
\bea 
& &V_{ud} = 0.97425\,, \quad V_{us} = 0.2252\,, \quad V_{ub} = 0.00389\,, 
\nonumber\\
& &V_{cd} = 0.230\,, \quad V_{cs} = 0.974642\,, \quad V_{cb} = 0.0406\,. 
\ena 
We are dealing with the quark level transitions (i) $b \to c \bar u d,\,                
b \to c \bar c s$, $c \to s \bar d u$  (CKM favored), (ii)
$b \to c \bar c d$, $b \to u \bar c s$, $b \to c \bar u s$,
$b \to u \bar u d$, $c \to s \bar s u$ (CKM singly suppressed), and
(iii) $b \to u \bar u s$, $b \to u \bar c d$ (CKM doubly suppressed).
\begin{enumerate}
\item
$b \to c \bar u q$ and $b \to u \bar c q$
transition with $q=d,s$
(CKM favored: $b \to c \bar u d$, CKM suppressed: $b \to c \bar u s$,
$b \to u \bar c s$, CKM doubly suppressed: $b \to u \bar c d$):
\bea
{\cal L}_{\rm eff} &=&
\frac{G_F}{\sqrt{2}} \, \biggl[
V^\dagger_{cb} \, V_{uq} \, (C_1 \, Q_1 \, + \, C_2 \, Q_2) \,+\,
V^\dagger_{ub} \, V_{cq} \, (C_1 \, \tilde Q_1 \, + \, C_2 \, \tilde Q_2)
\biggr]
\,+\, {\rm H.c.}
\ena
where $Q_i$ and $\tilde Q_i$
is the set of flavor-changing effective four-quark $b \to c \bar u d(s)$ and
$b \to u \bar c d(s)$ operators
\bea
& &    Q_1 = (\bar c^{a_1} O^\mu b^{a_2}) \, (\bar{q}^{a_2} O_\mu u^{a_1}) \,,
\quad  Q_2 = (\bar c^{a_1} O^\mu b^{a_1}) \, (\bar{q}^{a_2} O_\mu u^{a_2}) \,,
\nonumber\\
& &\tilde Q_1 = (\bar u^{a_1} O^\mu b^{a_2}) \, (\bar{q}^{a_2} O_\mu c^{a_1}) \,,
\quad
   \tilde Q_2 = (\bar u^{a_1} O^\mu b^{a_1}) \, (\bar{q}^{a_2} O_\mu c^{a_2})
\,,
\ena
$O^\mu = \gamma^\mu (1 - \gamma^5)$,
$C_i$ is the set of Wilson coefficients~\cite{Buchalla:1995vs,%
Altmannshofer:2008dz,Neubert:1991we}:
\be
   C_1 = -0.25\,, \quad
   C_2 = 1.1\,.
\label{eq:Wilson_bc}
\en
The quark-level matrix elements contributing
to the bottom-charm baryon and bottom-up baryon transitions
are given by
\be
M(b\to c \bar u q)  = \frac{G_F}{\sqrt{2}} \,
C_{\rm eff}^{(bc)} \, V^\dagger_{cb} \,V_{uq} \,
\left( \bar c \, O^\mu \, b \right)  \,
\left( \bar q \, O_\mu \, u \right) \,, \quad 
M(b\to q \bar u c)  = \frac{G_F}{\sqrt{2}} \,
C_{\rm eff}^{(bq)} \, V^\dagger_{cb} \,V_{uq} \,
\left( \bar q \, O^\mu \, b \right)  \,
\left( \bar c \, O_\mu \, u \right) \,  
\label{eq:free_quark_bcuq}
\en
and
\be
M(b\to u \bar c q)  = \frac{G_F}{\sqrt{2}} \,
\tilde C_{\rm eff}^{(bu)} \, V^\dagger_{ub} \,\, V_{cq}
\left( \bar u \, O^\mu \, b \right)  \,
\left( \bar q \, O_\mu \, c \right) \,, \quad 
M(b\to c \bar u q)  = \frac{G_F}{\sqrt{2}} \,
\tilde C_{\rm eff}^{(bc)} \, V^\dagger_{ub} \,\, V_{cq}
\left( \bar u \, O^\mu \, b \right)  \,
\left( \bar q \, O_\mu \, c \right) \,, 
\label{eq:free_quark_bucq}
\en
where
\be
C_{\rm eff}^{(bc)} = \tilde C_{\rm eff}^{(bu)} = 
\xi C_1 + C_2 \,, \quad 
C_{\rm eff}^{(bq)} = \tilde C_{\rm eff}^{(bq)} = 
C_1 + \xi C_2 \,. 
\label{eq:color-factor_bc_bu}
\en 

\item
$b \to c \bar c q$ transition with $q=d,s$
(CKM favored:   $b \to c \bar c s$,
CKM suppressed: $b \to c \bar c d$):
\be
{\cal L}_{\rm eff} =
V_{cb} \, V^\dagger_{cq} \, \sum\limits_{i=1}^6
\, C_i \, Q_i \,+\, {\rm H.c.} \,,
\label{eq:weak_Lag}
\en
where the $Q_i$ are the set of effective four-quark flavor-changing
$b \to q$
operators
\bea
Q_1 &=& (\bar c^{a_1} O^\mu b^{a_2}) \, (\bar q^{a_2} O_\mu c^{a_1}) \,,
\qquad
Q_4 = (\bar q^{a_1} O^\mu b^{a_2}) \, (\bar c^{a_2} O_\mu c^{a_1}) \,,
\nn
Q_2 &=& (\bar c^{a_1} O^\mu b^{a_1}) \, (\bar q^{a_2} O_\mu c^{a_2}) \,,
\qquad
Q_5 = (\bar s^{a_1} O^\mu b^{a_1}) \, (\bar c^{a_2} \tilde O_\mu c^{a_2}) \,,
\nn
Q_3 &=& (\bar q^{a_1} O^\mu b^{a_1}) \, (\bar c^{a_2} O_\mu c^{a_2}) \,,
\qquad
Q_6 = (\bar q^{a_1} O^\mu b^{a_2}) \, (\bar c^{a_2} \tilde O_\mu c^{a_1}) \,,
\label{eq:Q_i}
\ena
with
$\tilde O^\mu = \gamma^\mu (1 + \gamma^5)$. The respective
Wilson coefficients are~\cite{Altmannshofer:2008dz}:
\be
   C_1 = - 0.257\,, \quad
   C_2 = 1.009\,,   \quad
   C_3 = - 0.005\,, \quad
   C_4 = - 0.078\,, \quad
   C_5 \simeq  0\,, \quad
   C_6 = 0.001 \,.
\label{eq:Wilson_bscc}
\en
The quark-level matrix elements contributing
to the bottom-charm baryon and bottom-down(strange) baryon transitions
are given by
\be
M(b\to c \bar c q)  = \frac{G_F}{\sqrt{2}} \,
C_{\rm eff}^{(bc)}\, V_{cb} \, V^\dagger_{cq} \,
\left( \bar c \, O^\mu \, b \right)  \,
\left( \bar q \, O_\mu \, c \right) \,,
\label{eq:free_quark_bccs}
\en
\be
M(b\to q \bar c c)  = \frac{G_F}{\sqrt{2}} \,
C_{\rm eff}^{(bq)} \, V_{cb} \, V^\dagger_{cq} \,
\left( \bar q \, O^\mu \, b \right)  \,
\left( \bar c \, O_\mu \, c \right) \,,
\label{eq:free_quark_bscc}
\en
where
\be
C_{\rm eff}^{(bc)} = \xi \Big(C_1 + C_3 + C_5\Big)
+ \Big(C_2 + C_4 + C_6\Big) \,.
\label{eq:color-factor_bccc}
\en
and
\be
C_{\rm eff}^{(bq)} = C_1 + C_3 + C_5
+ \xi \Big(C_2 + C_4 + C_6\Big) \,.
\label{eq:color-factor_bscc}
\en
\item
$c \to s \bar q u$ transition with $q=d,s$
(CKM favored: $c \to s \bar d u$, CKM suppressed: $c \to s \bar s u$):
\be
{\cal L}_{\rm eff} =
\frac{G_F}{\sqrt{2}} \,
V^\dagger_{cs} \, V_{uq} \, (C_1 \, Q_1 \, + \, C_2 \, Q_2) \,+\, {\rm H.c.}
\en
$Q_i$ is the set of flavor-changing effective four-quark $c \to s$
operators
\be
Q_1 = (\bar s^{a_1} O^\mu c^{a_1}) \, (\bar u^{a_2} O_\mu q^{a_2}) \,,
\quad 
Q_2 = (\bar s^{a_1} O^\mu c^{a_2}) \, (\bar u^{a_2} O_\mu q^{a_1}) \,,
\en
where
\be
   C_1 = 1.26\,, \quad
   C_2 = -0.51\,.
\label{eq:Wilson_csuq}
\en
The quark-level matrix elements contributing
to the charm-to-nonstrange (e.g., $\Lambda_c \to p \phi$) and
charm-to-strange baryon decay (e.g., $\Lambda_c \to \Lambda \pi^+$) are given by
\be
M(c\to u \bar s s)  = \frac{G_F}{\sqrt{2}} \,
C_{\rm eff}^{(cu)}\, V^\dagger_{cs} \, V_{us} \,
\left( \bar u \, O_\mu \, c \right)  \,
\left( \bar s \, O^\mu \, s \right) \,,
\en
where
\be
C_{\rm eff}^{(cu)} = C_2 + \xi C_1
\label{eq:color-factor_cuss}
\en
and
\be
M(c\to s \bar d u)  = \frac{G_F}{\sqrt{2}} \,
C_{\rm eff}^{(cs)}\, V^\dagger_{cs} \, V_{ud} \,
\left( \bar s \, O_\mu \, c \right)  \,
\left( \bar u \, O^\mu \, d \right) \,,
\en
with
\be
C_{\rm eff}^{(cs)} = C_1 + \xi C_2 \,.
\label{eq:color-factor_csud}
\en

\item
$b \to u \bar u q$ transition (CKM singly suppressed: $b \to u \bar u d$,
doubly suppressed: $b \to u \bar u s$):
\be
{\cal L}_{\rm eff} =
\frac{G_F}{\sqrt{2}} \,
V^\dagger_{ub} \, V_{uq} \, (C_1 \, Q_1 \, + \, C_2 \, Q_2) \,+\, {\rm H.c.}
\en
$Q_i$ is the set of flavor-changing effective four-quark $b \to u$
operators
\be
Q_1 = (\bar u^{a_1} O^\mu b^{a_1}) \, (\bar q^{a_2} O_\mu u^{a_2}) \,,
\quad 
Q_2 = (\bar u^{a_1} O^\mu b^{a_2}) \, (\bar q^{a_2} O_\mu u^{a_1}) \,,
\en
here
\be
   C_1 = 1.26\,, \quad
   C_2 = -0.51\,.
\label{eq:Wilson_buus}
\en
The quark-level matrix elements contributing to the 
$\Xi_b^- \to \Sigma^- \pi^0(\eta,\eta',\rho^0,\omega)$
and $\Xi_b^- \to \Xi^- \pi^0(\eta,\eta',\rho^0,\omega)$
are given by
\be
M(b\to q \bar u u)  = \frac{G_F}{\sqrt{2}} \,
C_{\rm eff}^{(bq)} V^\dagger_{ub} \, V_{uq} \,
\left( \bar q \,  O_\mu \, b \right)  \,
\left( \bar u \,  O^\mu \, u \right) \,,
\en 
where
\be
C_{\rm eff}^{(bq)} = C_2 + \xi C_1 \;.
\label{eq:color-factor_buus}
\en
\end{enumerate}
The color factor $\xi=1/N_c$
will be set to zero in all matrix elements, i.e. we keep only the leading term in the
$1/N_c-$ expansion.

\section{Matrix elements, helicity amplitudes and rate expressions}

The matrix element of the exclusive decay 
$B_1(p_1,\lambda_1)\to B_2(p_2,\lambda_2)\,+\,M(q,\lambda_M)$ 
is defined by

\be
M(B_1\to B_2 + M) =  
\frac{G_F}{\sqrt{2}} \, V_{ij} \, V^\dagger_{kl} \, C_{\rm eff} \, 
f_M \, M_M \, \la B_2 | \bar q_2 O_\mu q_1 | B_1 \ra \, 
\epsilon^{\dagger\,\mu}(\lambda_M) \,,
\label{eq:matr_LbLJ}
\en 
where $M=V, P$ stands for the vector and pseudoscalar mesons.
$M_M$ and $f_M$ are the respective masses $M_V,\,M_P$
and leptonic decay constants $f_V,\,f_P$. The Dirac string $O^\mu$ is defined by
$O^\mu = \gamma^\mu (1 - \gamma^5)$.

The hadronic matrix element $\la B_2 | \bar q_2 O^\mu q_1 | B_1 \ra$
is expressed in terms of six ($1/2^+ \to 1/2^+$) or eight ($1/2^+ \to 3/2^+$),
dimensionless invariant form factors
$F^{V/A}_i(q^2)$, viz.
\begin{itemize}
\item
Transition $\frac{1}{2}^+ \to \frac{1}{2}^+$\,: 
\bea 
\la B_2 | \bar q_2 \gamma_\mu q_1 | B_1 \ra &=& 
\bar u(p_2,s_2) 
\Big[ \gamma_\mu F_1^V(q^2) 
    - i \sigma_{\mu\nu} \frac{q_\nu}{M_1} F_2^V(q^2)  
    + \frac{q_\mu}{M_1} F_3^V(q^2) 
\Big] 
u(p_1,s_1) 
\nn
\la B_2 |\bar q_2\gamma_\mu\gamma_5 q_1  | B_1 \ra &=& 
\bar u(p_2,s_2) 
\Big[ \gamma_\mu F_1^A(q^2)  
    - i \sigma_{\mu\nu} \frac{q_\nu}{M_1} F_2^A(q^2)   
    + \frac{q_\mu}{M_1} F_3^A(q^2)  
\Big] \gamma_5 u(p_1,s_1) 
\nonumber
\ena
 
\item
Transition $\frac{1}{2}^+ \to \frac{3}{2}^+$\,: 
\bea
\la B_2^\ast |\bar q_2 \gamma_\mu q_1| B_1 \ra &=& 
\bar u^\alpha(p_2,s_2) 
\Big[ g_{\alpha\mu} F_1^V(q^2)  
    + \gamma_\mu \frac{p_{1\alpha}}{M_1} F_2^V(q^2) 
+ \frac{p_{1\alpha} p_{2\mu}}{M_1^2} F_3^V(q^2)   
    + \frac{p_{1\alpha} q_\mu}{M_1^2}    F_4^V(q^2)       
\Big] \gamma_5 
u(p_1,s_1) 
\nn
\la B_2^\ast |\bar q_2 \gamma_\mu\gamma_5 q_1 | B_1 \ra &=& 
\bar u^\alpha(p_2,s_2) 
\Big[ g_{\alpha\mu} F_1^A(q^2)  
    + \gamma_\mu \frac{p_{1\alpha}}{M_1} F_2^A(q^2) 
+ \frac{p_{1\alpha} p_{2\mu}}{M_1^2} F_3^A(q^2)   
    + \frac{p_{1\alpha} q_\mu}{M_1^2}    F_4^A(q^2)       
\Big] 
u(p_1,s_1) 
\nonumber
\ena  
\end{itemize}
where 
$\sigma_{\mu\nu} = (i/2) (\gamma_\mu \gamma_\nu - \gamma_\nu \gamma_\mu)$ 
and all $\gamma$ matrices are defined as in Bjorken-Drell.  
We use the same notation for the form factors $F^{V/A}_i$ in all
transitions even though their numerical values are different. 

Next we express the vector and axial helicity amplitudes 
$H_{\lambda_2\lambda_M}$ in terms of the invariant form factors 
$F_i^{V/A}$,  where $\lambda_M = t, \pm 1, 0$ and 
$\lambda_2 = \pm 1/2, \pm 3/2$ are  the helicity components of the meson
$M\,(M=P,V)$ and the baryon $B_2$,  respectively. 
We need to calculate the expressions 
\be
H_{\lambda_2\lambda_M} = 
\la B_2(p_2,\lambda_2) |\bar q_2 O_\mu q_1 | B_1(p_1,\lambda_1) \ra
\epsilon^{\dagger\,\mu}(\lambda_M) 
=  H_{\lambda_2\lambda_M}^V - H_{\lambda_2\lambda_M}^A
\en 
where we split the  helicity amplitudes into their vector and axial parts.
For the color enhanced decays the operator $\bar q_2 O_\mu q_1$ represents a
charged current transition while the operator $\bar q_2 O_\mu q_1$ describes a
neutral current transition for the color suppressed decays.
We shall work in the rest frame of the baryon $B_1$ with the baryon $B_2$ 
moving in the positive $z$-direction:
$p_1 = (M_1, \vec{\bf 0})$, $p_2 = (E_2, 0, 0, |{\bf p}_2|)$ and 
$q = (q_0, 0, 0, - |{\bf p}_2|)$. The helicities of the three particles
are related by $\lambda_1 = \lambda_2 - \lambda_M$. We use the notation
$\lambda_P=\lambda_t=0$ for the scalar $(J=0)$ contribution to set the
helicity label apart from
$\lambda_V=0$ used for the longitudinal compo<nent of the $J=1$ vector meson.
One has

\vspace*{.15cm} 
\begin{itemize}
\item \noindent
Transition $\frac12^+ \to \frac12^+$\,:
$H^V_{-\lambda_2,-\lambda_M} = + H^V_{\lambda_2,\lambda_M}$ and 
$H^A_{-\lambda_2,-\lambda_M} = - H^A_{\lambda_2,\lambda_M}$\,. 

\[
\begin{array}{lcrlcl}
H_{\frac12 t}^V &=& \sqrt{Q_+/q^2} \, 
\Big( F_1^V M_- + F_3^V \frac{q^2}{M_1} \Big) \qquad 
&\qquad 
H_{\frac12 t}^A &=& \sqrt{Q_-/q^2} \, 
\Big( F_1^A M_+ - F_3^A \frac{q^2}{M_1} \Big) 
\\[1.1ex]
H_{\frac12 0}^V &=& \sqrt{Q_-/q^2} \,
\Big( F_1^V M_+ + F_2^V \frac{q^2}{M_1} \Big) \qquad 
& \qquad 
H_{\frac12 0}^A &=& \sqrt{Q_+/q^2} \,   
\Big( F_1^A M_- - F_2^A \frac{q^2}{M_1} \Big) 
\\[1.1ex]
H_{\frac12 1}^V &=& \sqrt{2Q_-} 
\Big( - F_1^V - F_2^V \frac{M_+}{M_1} \Big) \qquad 
& \qquad 
H_{\frac12 1}^A &=& \sqrt{2Q_+} \,  
\Big( - F_1^A + F_2^A \frac{M_-}{M_1} \Big) 
\\
\end{array}
\]

\item
Transition $\frac12^+ \to \frac32^+$\,: 
$H^V_{-\lambda_2,-\lambda_M} = - H^V_{\lambda_2,\lambda_M}$ and 
$H^A_{-\lambda_2,-\lambda_M} = +  H^A_{\lambda_2,\lambda_M}$. 

\bea 
H_{\frac12 t}^V &=& - \sqrt{\frac23\cdot \frac{Q_+}{q^2}} \, 
\frac{Q_-}{2M_1M_2} 
\Big( F_1^V M_1 - F_2^V M_+ + F_3^V \frac{M_+M_--q^2}{2M_1}  
+ F_4^V \frac{q^2}{M_1} \Big) 
\nn[1.1ex]
H_{\frac12 0}^V &=& - \sqrt{\frac23\cdot \frac{Q_-}{q^2}} \, 
\Big( F_1^V \frac{M_+M_--q^2}{2M_2} - F_2^V \frac{Q_+ M_-}{2M_1M_2}  
+ F_3^V \frac{|{\bf p_2}|^2}{M_2} \Big) 
\nn[1.1ex]
H_{\frac12 1}^V &=& \sqrt{\frac{Q_-}{3}} \,  
\Big( F_1^V - F_2^V \frac{Q_+}{M_1M_2} \Big) 
\qquad
H_{\frac32 1}^V = -  \, \sqrt{Q_-} \, F_1^V 
\nn[2ex]
H_{\frac12 t}^A &=& \sqrt{\frac23\cdot \frac{Q_-}{q^2} }
\frac{Q_+}{2M_1M_2} 
 \Big( F_1^A M_1 + F_2^A M_- + F_3^A \frac{M_+M_--q^2}{2M_1}   
+ F_4^A \frac{q^2}{M_1}\Big) 
\nn[1.1ex]
H_{\frac12 0}^A &=&  \sqrt{\frac23\cdot \frac{Q_+}{q^2} } 
\Big( F_1^A \frac{M_+M_--q^2}{2M_2} + F_2^A \frac{Q_-M_+}{2M_1M_2} 
+ F_3^A  \frac{|{\bf p_2}|^2}{M_2}  \Big) 
\nn[1.1ex]
H_{\frac12 1}^A &=& \sqrt{\frac{Q_+}{3}} 
\Big( F_1^A - F_2^A \frac{Q_-}{M_1M_2} \Big) 
\qquad
H_{\frac{3}{2}1}^A = \, \sqrt{Q_+} F_1^A 
\nonumber
\ena 
\end{itemize}

\noindent We use the abbreviations
$M_\pm = M_1 \pm M_2$, 
$Q_\pm = M_\pm^2 - q^2$,
${|\bf p_2|} = \lambda^{1/2}(M_1^2,M_2^2,q^2)/(2M_1)$. 
\vspace{0.2cm}

\noindent For the decay width one finds 
\be
\Gamma(B_1 \to B_2\,+\,M) 
= \frac{G_F^2}{32 \pi} \, \frac{|{\bf p_2}|}{M_1^2} \, 
|V_{ij} V^\dagger_{kl}|^2 \, C_{\rm eff}^2 \, f_M^2 \, M_M^2 \,{\cal H}_N  
\label{eq:LbLV_width}
\en
where we denote the sum of the squared moduli of the  helicity amplitudes
by  
\be
   {\cal H}_N= \sum_{\lambda_2,\lambda_M}|H_{\lambda_2,\lambda_M}|^2.
   \en
   We have chosen to label the sum of the squared moduli of the  helicity
   amplitudes by the subscript $N$ since ${\cal H}_N$ is the common divisor
   needed for the normalization of the spin density matrix elements to be
   discussed in Sec.~VII. The sum over the helicities of the daughter baryon
   runs over $\lambda_2=\pm 1/2$ and $\lambda_2=\pm 1/2, \pm 3/2$ for the
($1/2^+ \to 1/2^+$) and ($1/2^+ \to 3/2^+$) transitions, respectively.
   For the vector meson case $M=V$ one sums over $\lambda_V=0,\pm 1$
   and for the pseudoscalar meson case $M=P$ one has the single value
   $\lambda_P=0$. Angular momentum conservation dictates the constraint
   $|\lambda_2 -\lambda_M|\le 1/2$.

\section{Interpolating three-quark currents of baryons and quantum numbers} 

We use the following generic notation for the baryonic interpolating 
currents: 
\begin{enumerate}
\item \quad Spin-parity $J^P = \frac{1}{2}^+$:
\be\label{Baryon_current_12p} 
J = \varepsilon^{a_1a_2a_3} J^{a_1a_2a_3}\,, \quad 
J^{a_1a_2a_3} = \Gamma_1 q^{a_1}_{m_1}  q^{a_2}_{m_2} C \Gamma_2 q^{a_3}_{m_3}\,,
\en 

\item \quad Spin-parity $J^P = \frac{3}{2}^+$:
\be\label{Baryon_current_32p}  
J_\mu = \varepsilon^{a_1a_2a_3} \, J^{a_1a_2a_3}_\mu\,, \quad 
J^{a_1a_2a_3}_\mu = q^{a_1}_{m_1}  q^{a_2}_{m_2} C \gamma_\mu q^{a_3}_{m_3}
\en
\end{enumerate}
where $\Gamma_1$ and $\Gamma_2$ are Dirac strings; 
$a_i$ and $m_i$ are color and flavor indices, respectively. 
The interpolating currents and quantum numbers of the baryons needed in this
paper are summarized in Table I. 
As indicated in the first entry of Table I we use a superposition of
two possible currents (vector and tensor) for the proton as 
motivated by nucleon phenomenology.

\begin{figure}
\begin{center}
\epsfig{figure=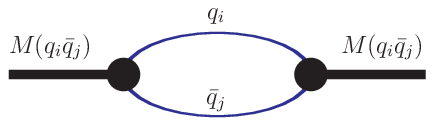,scale=.8} 
\caption{Diagram representing  the mass operator of a $M(q_i\bar q_j)$ meson.} 
\label{fig:Mass_M}
\end{center}

\vspace*{.5cm}

\begin{center}
\epsfig{figure=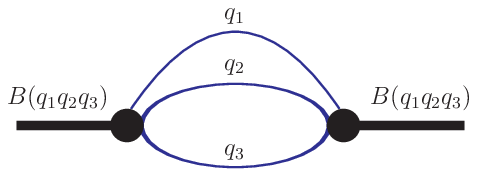,scale=.8} 
\caption{Diagram representing  the mass operator of a $B(q_1q_2q_3)$ baryon.} 
\label{fig:Mass_B}
\end{center}

\vspace*{.5cm}

\begin{center}
\epsfig{figure=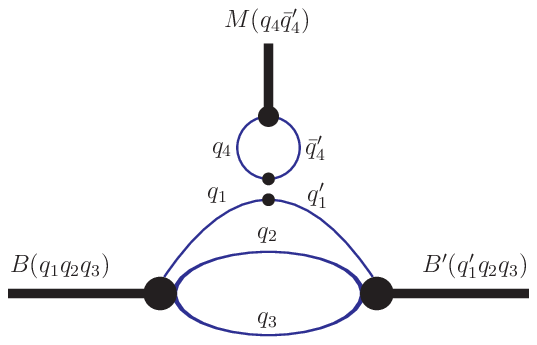,scale=.8} 
\end{center}
\caption{Factorizing diagram describing the $B(q_1q_2q_3) \to B'(q_1'q_2q_3) + M(q_4\bar q_4')$ 
transition.}
\label{fig:Decay} 
\end{figure}

The Lagrangian describing the coupling of a baryon with its constituent quarks 
is constructed using a nonlocal extension of the interpolating currents in 
Eqs.~(\ref{Baryon_current_12p}) and 
(\ref{Baryon_current_32p}) (see details in 
Refs.~\cite{Ivanov:1996pz}-\cite{Faessler:2001mr},%
\cite{Gutsche:2013pp,Ivanov:1997ra}).  
In particular, the interaction Lagrangian for the $\frac{1}{2}^+$ 
baryon $B(q_1q_2q_3)$ composed of constituent quarks $q_1, q_2$, and 
$q_3$ is written as 
\bea
B(q_1q_2q_3):\qquad &&{\cal L}^B_{\rm int}(x)
 = g_B \,\bar B(x)\cdot J_B(x) 
+ \mathrm{H.c.}\,,
\nonumber\\
&&J_B(x)
= \int\!\! dx_1 \!\! \int\!\! dx_2 \!\! \int\!\! dx_3 \, F_B(x;x_1,x_2,x_3) \,
\varepsilon^{a_1a_2a_3} \, \Gamma_1 \,
q^{a_1}_{m_1}(x_1)\,q^{a_2}_{m_2}(x_2) \,C \, \Gamma_2 \,
q^{a_3}_{m_3}(x_3)\,,
\ena 
We treat each of the three constituent quarks as separate dynamic entities,
i.e. we do not use the quark-diquark approximation for baryons
as done in many applications.
The propagators $S_q(k) = 1/(m_q \, - \not\!\! k)$ for quarks of all flavors 
$q=u,d,s,c,b$ are taken in the form of free fermion propagators.
The constituent quark masses $m_q$ were fixed
in a previous analysis of a multitude of hadronic processes
(see, e.g., Refs.~\cite{Gutsche:2013oea,Gutsche:2017wag}). 
By analogy we treat mesons as bound states of constituent quarks, i.e. 
we construct respective nonlocal Lagrangians for the mesons interacting 
with their constituent quarks (see details in 
Refs.~\cite{Branz:2009cd}-\cite{Dubnicka:2013vm}).   
The compositeness condition of Salam and Weinberg~\cite{Salam1962}
gives a constraint equation between the coupling
factors $g_B$ and the size parameters
$\Lambda_B$ 
characterizing the nonlocal distribution
$F_B(x;x_1,x_2,x_3)$. The constraint equation follows from the calculation of
the mass 
operators for mesons (Fig.~\ref{fig:Mass_M}) and baryons
(Fig.~\ref{fig:Mass_B}),   
As size parameters we use the values listed in Sec.~VI 
for the nucleon, light hyperons, single charm and single heavy baryons, 
respectively. 

\begin{table}[htb]
\begin{center}
\caption{Interpolating currents
and quantum numbers of baryons}

\vspace*{.15cm}

\def\arraystretch{1.5}
\begin{tabular}{c|c|c|c}
\hline
\ \ Baryon \ \ & \ \ $J^P$ \ \ & \ \ $J^{abc}$ \ \
& \ \ Mass (MeV) \ \ \\
\hline
$p$ & $\frac{1}{2}^+$ & $(1-x_N) \gamma^\mu\gamma^5 d^a u^b C\gamma_\mu u^c$ & 938.27 \\
\
  &                 & $+ x_N \sigma^{\mu\nu}\gamma^5 d^a u^b C\sigma_{\mu\nu} u^c$\,,
\ $x_N = 0.8$ & \\
$n$ & $\frac{1}{2}^+$ & $(1-x_N) \gamma^\mu\gamma^5 u^a d^b C\gamma_\mu d^c$ & 939.57 \\
\
  &                 & $+ x_N \sigma^{\mu\nu}\gamma^5 u^a d^b C\sigma_{\mu\nu} d^c$\,,
\ $x_N = 0.8$ & \\
\hline
$\Lambda$ & $\frac{1}{2}^+$ & $s^a u^b C\gamma_5 d^c$ & 1115.68 \\
\hline
$\Sigma^{+}$ & $\frac{1}{2}^+$ &
$\gamma^\mu\gamma^5 s^a u^b C\gamma_\mu u^c$
& 1189.37 \\
\hline
$\Sigma^{0}$ & $\frac{1}{2}^+$ &
$\sqrt{2} \gamma^\mu\gamma^5 s^a u^b C\gamma_\mu d^c$
& 1192.64 \\
\hline
$\Sigma^{-}$ & $\frac{1}{2}^+$ &
$\gamma^\mu\gamma^5 s^a d^b C\gamma_\mu d^c$
& 1197.45 \\
\hline
$\Xi^0$   & $\frac{1}{2}^+$ & $\gamma^\mu\gamma^5 u^a s^b C\gamma_\mu s^c$ & 1314.86 \\
\hline
$\Xi^-$   & $\frac{1}{2}^+$ & $\gamma^\mu\gamma^5 d^a s^b C\gamma_\mu s^c$ & 1321.71 \\
\hline
$\Sigma^{* +}$ & $\frac{3}{2}^+$ & $\frac{1}{\sqrt{3}} 
(s^a u^b C\gamma_\mu u^c + 2 u^a u^b C\gamma_\mu s^c)$
& 1382.80 \\
\hline
$\Sigma^{* 0}$ & $\frac{3}{2}^+$ & $\sqrt{\frac{2}{3}}                                 
(s^a u^b C\gamma_\mu d^c + u^a d^b C\gamma_\mu s^c + d^a u^b C\gamma_\mu s^c)$
& 1383.70 \\
\hline
$\Sigma^{* -}$ & $\frac{3}{2}^+$ & $\frac{1}{\sqrt{3}}                                 
(s^a d^b C\gamma_\mu d^c + 2 d^a d^b C\gamma_\mu s^c)$
& 1387.20 \\
\hline
$\Xi^{* 0}$ & $\frac{3}{2}^+$ &
$\frac{1}{\sqrt{3}}                                                                    
(u^a s^b C\gamma_\mu s^c + 2 s^a s^b C\gamma_\mu u^c)$
& 1531.80 \\
\hline
$\Xi^{* -}$ & $\frac{3}{2}^+$ & 
$\frac{1}{\sqrt{3}}                                                                     
(d^a s^b C\gamma_\mu s^c + 2 s^a s^b C\gamma_\mu d^c)$
& 1535.00 \\
\hline
$\Omega^-$  & $\frac{3}{2}^+$ & $s^a s^b C\gamma_\mu s^c$ & 1672.45 \\
\hline
$\Lambda_c$ & $\frac{1}{2}^+$ & $ c^a u^b C\gamma_5 d^c$ & 2286.46 \\
\hline
$\Xi_c^+$ & $\frac{1}{2}^+$ & $ c^a u^b C\gamma_5 s^c$ & 2467.87 \\
\hline
$\Xi_c^0$ & $\frac{1}{2}^+$ & $ c^a d^b C\gamma_5 s^c$ & 2470.87 \\
\hline
$\Xi_c^{'+}$ & $\frac{1}{2}^+$ & $ \gamma^\mu\gamma^5 c^a u^b C\gamma_\mu s^c$ 
& 2577.40 \\
\hline
$\Xi_c^{'0}$ & $\frac{1}{2}^+$ & $ \gamma^\mu\gamma^5 c^a d^b C\gamma_\mu s^c$ 
& 2578.80 \\
\hline
$\Omega_c$  & $\frac{1}{2}^+$ & $ \gamma^\mu\gamma^5 c^a s^b C\gamma_\mu s^c$ 
& 2695.20 \\
\hline
$\Omega_c^*$ & $\frac{3}{2}^+$ & $ c^a s^b C\gamma_\mu s^c$ & 2765.90 \\
\hline
$\Lambda_b$      & $\frac{1}{2}^+$ & $ b^a u^b C\gamma_5 d^c$   & 5619.40 \\
\hline
$\Xi_b^0$        & $\frac{1}{2}^+$ & $ b^a u^b C\gamma_5 s^c$   & 5791.90 \\
\hline
$\Xi_b^-$        & $\frac{1}{2}^+$ & $ b^a d^b C\gamma_5 s^c$   & 5794.50 \\
\hline
$\Xi_b^{'0}$     & $\frac{1}{2}^+$ & $ \gamma^\mu\gamma^5 b^a u^b C\gamma_\mu s^c$ 
& 5935.02 \\
\hline
$\Xi_b^{'-}$      & $\frac{1}{2}^+$ & $ \gamma^\mu\gamma^5 b^a d^b C\gamma_\mu s^c$ 
& 5935.02 \\
\hline
$\Omega_b$       & $\frac{1}{2}^+$ & $ \gamma^\mu\gamma^5 b^a u^b C\gamma_\mu s^c$ 
& 6046.10 \\
\hline
\end{tabular}
\label{tab:baryons}
\end{center}
\end{table}

\clearpage 

\section{Model Parameters}

Our covariant constituent quark model (CCQM)~\cite{Branz:2009cd}-\cite{Gutsche:2013pp}  
contains a number of model parameters which have been determined by a global fit to 
a multitude of decay processes. The best fit values for the constituent quark masses 
and the universal infrared cutoff parameter are given by
\be
\def\arraystretch{2}
\begin{array}{ccccccc}
     m_u        &      m_s        &      m_c       &     m_b & \lambda  &
\\\hline
 \ \ 0.242\ \   &  \ \ 0.428\ \   &  \ \ 1.672\ \   &  \ \ 5.046\ \   &
\ \ 0.181\ \   & \ {\rm GeV}
\end{array}
\label{eq: fitmas}
\en
The nonlocal quark particle vertices are described in terms of a number of
hadronic scale or size parameters $\Lambda_H$ (in GeV) for which the best fit
values are
\be
\def\arraystretch{2}
\begin{array}{cccccccc}
\hline\hline
  \pi  & \eta(\Lambda_\eta^{q \bar q}, \Lambda_\eta^{s \bar s}) 
       & \eta'(\Lambda_{\eta'}^{q \bar q}, \Lambda_{\eta'}^{s \bar s}) 
       & \rho  & \omega & \phi  & D   & D^*  \\  
\hline
   \ \ 0.871\ \   
&  \ \ (0.881,1.973)\ \
&  \ \ (0.257,2.797)\ \
&  \ \ 0.610\ \   
&  \ \ 0.490\ \   
&  \ \ 0.880\ \   
&  \ \ 1.600\ \   
&  \ \ 1.530\ \   
\\
\hline
\hline
  D_s  & D_s^* &  B    & B^*   & B_s  & B_s^* & \eta_c &  J/\psi \\ 
\hline
   \ \ 1.750\ \   
&  \ \ 1.560\ \ 
&  \ \ 1.960\ \   
&  \ \ 1.710\ \  
&  \ \ 2.050\ \   
&  \ \ 1.710\ \   
&  \ \ 2.200\ \ 
&  \ \ 1.738\ \  
\\
\hline
\hline
p & \Lambda & \Sigma & \Xi & \Sigma^* & \Xi^* & \Omega & \\
\hline

\ \ 0.3600 \ \ &  
\ \ 0.4925\ \ &  
\ \ 0.4925\ \ &  
\ \ 0.4925\ \ &  
\ \ 0.3201\ \ &  
\ \ 0.3201\ \ &  
\ \ 0.3201\ \ &  
\\
\hline
\hline
\Lambda_c & \Xi_c & \Omega_c & \Omega_c^* & \Lambda_b & \Xi_b  & \Omega_b & \\
\ \ 0.8675\ \ & 
\ \ 0.8675\ \ &  
\ \ 0.8675\ \ &  
\ \ 0.8675\ \ &  
\ \ 0.5709\ \ &  
\ \ 0.5709\ \ &  
\ \ 0.5709\ \ &
\\ 
\hline
\hline
\end{array}
\label{eq: fitLambda}
\en
Note, in the case of pseudoscalar mesons we introduce singlet-octet mixing
with a mixing angle of $\theta_{P} = -13.34^\circ$ deduced 
from Ref.~\cite{Ambrosino:2006gk} (see details 
in Refs.~\cite{Branz:2009cd,Dubnicka:2013vm}) 
while the vector mesons are assumed to be ideally mixed: 
\bea 
\eta &\longrightarrow&
-\,\tfrac{1}{\sqrt{2}}\sin\delta\, \Phi_{\Lambda_{\eta}}(\bar u u +\bar d d)
-\,\cos\delta\, \Phi_{\Lambda_{\eta_s}}\,\bar s s \,,
\nonumber\\
\eta' &\longrightarrow&
+\,\tfrac{1}{\sqrt{2}}\cos\delta\, \Phi_{\Lambda_{\eta'}}\,(\bar u u +\bar
d d)
-\,\sin\delta\, \Phi_{\Lambda_{\eta'_s}}\, \bar s s \,,
\nonumber\\
\delta &=& \theta_P-\theta_I, \qquad \theta_I=\arctan{\tfrac{1}{\sqrt{2}}}.
\label{eq:mixing}
\ena
We used~\cite{Branz:2009cd,Dubnicka:2013vm} 
the experimental data of the electromagnetic decays
involving $\eta$ and $\eta'$ to fit the size parameters 
defining the distribution of nonstrange ($q = u,d$) and 
strange $s$ quarks in these states: 
$\Lambda_\eta^{q\bar q}$, $\Lambda_{\eta'}^{q\bar q}$, 
$\Lambda_\eta^{s\bar s}$, and $\Lambda_{\eta'}^{s\bar s}$. 

An important input to the global fit are the mesonic decay constants.
Our global fit results in values for the mesonic decay constants which are 
quite close to the experimental decay constants~\cite{PDG:2016},
results of lattice QCD calculations~\cite{Becirevic:1998ua,Becirevic:2013bsa} 
and of nonrelativistic QCD~\cite{Braguta:2006wr} (see Table~\ref{tab:fH}).  

\begin{table}[t!]
\begin{center}
\caption{Results for the leptonic decay constants $f_H$ (in MeV).} 

\vspace*{.15cm}

\def\arraystretch{1.25}
\begin{tabular}{l|c|l}
  \hline
$f_H$ & Our results & Data/Lattice QCD \\
  \hline
$f_{\pi^+}$ & 130.3 & 130.2 $\pm$ 1.7~\cite{PDG:2016} \\
$f_{\rho^+}$    & 218.3 & 221 $\pm$ 1~\cite{PDG:2016} \\
$f_\omega$      & 198.7 & 198 $\pm$ 2~\cite{PDG:2016} \\
$f_\phi$        & 226.6 & 227 $\pm$ 2~\cite{PDG:2016} \\
$f_D$           & 204.7 & 203.7 $\pm$ 4.7 $\pm$ 0.6~\cite{PDG:2016} \\
$f_{D^*}$   & 244.4 & $245 \pm 20^{+3}_{-2}$~\cite{Becirevic:1998ua} \\
$f_{D_s}$       & 256.4 & 257.8 $\pm$ 4.1 $\pm$ 0.1~\cite{PDG:2016} \\
$f_{D_s^*}$ & 272.6 & $245 \pm 20^{+3}_{-2}$~\cite{Becirevic:1998ua} \\
$f_{\eta_c}$    & 426.8 & 387 $\pm$ 7 $\pm$ 2~\cite{Becirevic:2013bsa} \\
                &       & 346 $\pm$ 17~\cite{Braguta:2006wr} \\
$f_{J/\psi}$    & 415.1 & 416 $\pm$ 5~\cite{PDG:2016} \\
$f_B$           & 192.3 & 188 $\pm$ 17 $\pm$ 18~\cite{PDG:2016} \\
$f_{B^*}$   & 184.6 & $196 \pm 24^{+39}_{-2}$~\cite{Becirevic:1998ua} \\
$f_{B_s}$       & 238.2 & 227.2 $\pm$ 3.4~\cite{PDG:2016} \\
$f_{B_s^*}$ & 217.1 & 229 $\pm$ 46~\cite{Becirevic:1998ua} \\
\hline
\end{tabular}
\label{tab:fH}
\end{center}
\end{table}

\section{Numerical results}

Using the numerical values for the model parameters listed 
in Sec.~VI we present our predictions for the branching ratios, helicity 
amplitudes, and asymmetry parameters  
of the nonleptonic decays of single heavy baryons 
in Tables~\ref{tab:rates_bccs}-\ref{tab:Lcpleptons_rates}. All nonleptonic 
decays are described by the generic 
tree diagram shown in Fig.~\ref{fig:Decay}. 
In our estimate for the branching ratio
${\rm Br}(\Lambda_b \to \Lambda J/\psi) = 
(8.3 \pm 1.1) \times 10^{-4}$ we used the fragmentation fraction of 
$b$ quark into $\Lambda$ baryon of $f_{\Lambda_b} = 7\%$ 
from~\cite{Galanti:2015pqa}. 
   
\begin{table}

\begin{center}
\caption{Branching ratios of nonleptonic two-body decays of heavy baryons 
(in units of $10^{-4}$): $b \to c;\,\,  c \to s$ quark-level transition.} 

\vspace*{.15cm}

\def\arraystretch{1.1}
\begin{tabular}{l|c|c|l}
\hline
Mode & Our results & Data & Theory \\
\hline
$\Lambda_b^0 \to \Lambda_c^+\,D_s^-$         
& 147.8  
& 110 $\pm$ 10
& $230^{+30}_{-40}$~\cite{Mannel:1992ti};
                110~\cite{Cheng:1996cs};
                223~\cite{Fayyazuddin:1998ap};
                 77~\cite{Mohanta:1998iu};
              129.1~\cite{Giri:1997te}
\\
$\Lambda_b^0 \to \Lambda_c^+\,D_s^{*\,-}$ 
& 251.6  &
& $173^{+20}_{-30}$~\cite{Mannel:1992ti};
                 91~\cite{Cheng:1996cs};
                326~\cite{Fayyazuddin:1998ap};
              141.4~\cite{Mohanta:1998iu};
              198.3~\cite{Giri:1997te}
\\
$\Lambda_b^0 \to \Lambda^0 \, \eta_c$           
& 4.3 &  &
1.5 $\pm$ 0.9~\cite{Hsiao:2015cda}\\
$\Lambda_b^0 \to \Lambda^0 J/\psi$           
& 8.3 & 8.3 $\pm$ 1.1
& 1.6~\cite{Cheng:1996cs};
  2.7~\cite{Ivanov:1997ra};
  6.0~\cite{Fayyazuddin:1998ap};
  2.5~\cite{Mohanta:1998iu};
  2.1~\cite{Cheng:1995fe};
  3.5 $\pm$ 1.8~\cite{Chou:2001bn}; \\
& & &
  3.3 $\pm$ 2.0~\cite{Hsiao:2015cda}
  7.8~\cite{Fayyazuddin:2017sxq}
  8.4~\cite{Wei:2009np};
  8.2~\cite{Mott:2011cx} 
\\
$\Xi_b^-   \to \Xi^- \eta_c$              & 1.7  & &
2.3 $\pm$ 1.4~\cite{Hsiao:2015cda}
\\
$\Xi_b^0   \to \Xi^0 \, \eta_c$              & 1.6  & &
\\
$\Xi_b^- \to \Xi^- J/\psi$                & 4.6  & &
4.9 $\pm$ 3.0~\cite{Hsiao:2015cda}
\\
$\Xi_b^0 \to \Xi^0 \, J/\psi$                & 4.4  & &
\\
$\Omega_b^-  \to \Omega^- \eta_c$         & 1.9  & &  
\\
$\Omega_b^-  \to \Omega^- J/\psi$         & 8.1  & &  
\\
\hline
\end{tabular}
\label{tab:rates_bccs}
\end{center}

\begin{center}
\caption{Branching ratios of nonleptonic two-body decays of heavy baryons 
(in units of $10^{-4}$): $b \to c; \,\, u \to d$ quark-level transition.} 

\vspace*{.15cm}

\def\arraystretch{1.1}
\begin{tabular}{l|c|c}
\hline
Mode & Our results & Theory \\
\hline
$\Omega_b^- \to \Omega_c^0 \, \pi^-$          &  18.8 & 58.1~\cite{Ivanov:1997ra} \\
$\Omega_b^- \to \Omega_c^0 \, \rho^-$         &  54.3 & \\
$\Omega_b^- \to \Omega_c^{\ast 0} \pi^-$      &  17.0 & \\
$\Omega_b^- \to \Omega_c^{\ast 0} \rho^-$     &  55.8 & \\
$\Xi_b^- \to \Sigma^- D^{0}$                  &   0.1 & \\
$\Xi_b^- \to \Sigma^- D^{\ast 0}$             &   0.3 & \\
$\Omega_b^- \to \Xi^- D^{0}$                  & 1.4 $\times 10^{-2}$ & \\
$\Omega_b^- \to \Xi^- D^{\ast 0}$             & 2.6 $\times 10^{-2}$ & \\
$\Omega_b^- \to \Xi^{\ast -}  D^{0}$          & 0.7 $\times 10^{-3}$ & \\
$\Omega_b^- \to \Xi^{\ast -}  D^{\ast 0}$     & 1.9 $\times 10^{-3}$ & \\
\hline
\end{tabular}
\label{tab:rates_bcud}
\end{center}

\begin{center}

\caption{Branching ratios of nonleptonic two-body decays of heavy baryons 
(in units of $10^{-4}$): $b \to c;\,\, c \to d$ quark-level transition.} 

\vspace*{.15cm}
  \def\arraystretch{1.1}
\begin{tabular}{l|c}
\hline
Mode & Our results \\
\hline
$\Xi_b^0 \to \Xi_c^+\,D^-$                   & 4.5  \\
$\Xi_b^0 \to \Xi_c^+\,D^{*\,-}$              & 9.5  \\
$\Omega_b^- \to \Omega_c^0 \, D^-$           & 2.4  \\
$\Omega_b^- \to \Omega_c^0\,D^{*\,-}$        & 3.0  \\
$\Omega_b^- \to \Omega_c^{*\,0} \, D^-$      & 1.6  \\
$\Omega_b^- \to \Omega_c^{*\,0} \, D^{*\,-}$ & 5.8  \\
$\Lambda_b^0 \to n \,\eta_c$                 & 0.2  \\
$\Lambda_b^0 \to n \,J/\psi$                 & 0.4  \\
$\Xi_b^- \to \Sigma^- \,\eta_c$              & 0.1  \\
$\Xi_b^- \to \Sigma^- \,J/\psi$              & 0.2 \\
$\Xi_b^0 \to \Lambda^0 \, \eta_c$            & $1.6 \times 10^{-2}$ \\
$\Xi_b^0 \to \Sigma^0 \, \eta_c$             & $2.9 \times 10^{-2}$ \\ 
$\Xi_b^0 \to \Lambda^0 \, J/\psi$            & $3.1 \times 10^{-2}$ \\ 
$\Xi_b^0 \to \Sigma^0 \, J/\psi$             & $7.0 \times 10^{-2}$ \\ 
$\Omega_b^-  \to \Xi^- \,\eta_c$             & $1.3 \times 10^{-2}$ \\ 
$\Omega_b^-  \to \Xi^- \,J/\psi$             & $1.8 \times 10^{-2}$ \\  
$\Omega_b^-  \to \Xi^{*\,-} \,\eta_c$        & $6.9 \times 10^{-2}$ \\  
$\Omega_b^-  \to \Xi^{*\,-} \,J/\psi$        & 0.2 \\
\hline
\end{tabular}
\label{tab:rates_bccd}
\end{center}

\end{table}

\begin{table}
  
  \begin{center}
\caption{Branching ratios of nonleptonic two-body decays of heavy baryons 
(in units of $10^{-6}$): 
$b \to c; \,\, u \to s$ quark-level transition.} 

\vspace*{.15cm}

\def\arraystretch{1.2}
\begin{tabular}{l|c}
\hline
Mode & Our results \\
\hline
$\Xi_b^-     \to \Xi^- D^0$            & 1.4  \\
$\Xi_b^-     \to \Xi^- D^{* 0}$        & 2.1  \\
$\Omega_b^-  \to \Omega^- D^0$         & 0.7  \\
$\Omega_b^-  \to \Omega^- D^{* 0}$     & 2.0  \\
\hline
\end{tabular}
\label{tab:rates_bcus}
  \end{center}

\begin{center}
\caption{Branching ratios of nonleptonic two-body decays of heavy baryons 
(in units of $10^{-6}$): 
$b \to u;\,\, c \to s$ quark-level transition.} 

\vspace*{.15cm}

\def\arraystretch{1.2}
\begin{tabular}{l|c|c|l}
\hline
Mode & Our results & Data & Theory \\
\hline
$\Lambda_b \to p \, D_s^-$                  
& 13.2 & $<$ 480 & 13.6~\cite{Wei:2009np}; $18 \pm 3$~\cite{Hsiao:2015cda}  
\\
$\Lambda_b \to p \, D_s^{* -}$ & 22.1 & & 6.7~\cite{Wei:2009np}; 
$8.8 \pm 2.2$~\cite{Hsiao:2015cda}     
\\
$\Omega_b^-  \to \Xi^{*\,0} D_s^-$              &  4.9  & & \\
$\Omega_b^-  \to \Xi^{*\,0} D_s^{* -}$          & 11.3  & & \\
$\Omega_b^-  \to \Omega^- \bar D^{\,0}$         &  0.1  & & \\
$\Omega_b^-  \to \Omega^- \bar D^{* \,0}$       &  0.3  & & \\
\hline
\end{tabular}
\label{tab:rates_bucs}
\end{center}

\begin{center}
\caption{Branching ratios of nonleptonic two-body decays of heavy baryons 
(in units of $10^{-8}$): 
$b \to u; \,\, u \to d$ quark-level transition.} 

\vspace*{.15cm}

\def\arraystretch{1.2}
\begin{tabular}{l|c}
\hline
Mode & Our results \\
\hline
$\Xi_b^-\to \Sigma^- \pi^0$            &  6.2  \\
$\Xi_b^-\to \Sigma^- \eta$             &  4.9  \\ 
$\Xi_b^-\to \Sigma^- \eta'$            & 15.2  \\
$\Xi_b^-\to \Sigma^- \rho^0$           & 19.0  \\
$\Xi_b^-\to \Sigma^- \omega$           & 15.8  \\
$\Omega_b^-\to \Xi^- \pi^0$            &  0.5  \\
$\Omega_b^-\to \Xi^- \eta$             &  0.4  \\
$\Omega_b^-\to \Xi^- \eta'$            &  1.4  \\
$\Omega_b^-\to \Xi^- \rho^0$           &  1.7  \\
$\Omega_b^-\to \Xi^- \omega$           &  1.4  \\ 
$\Omega_b^-\to \Xi^{*\,-} \pi^0$       &  2.3  \\
$\Omega_b^-\to \Xi^{*\,-} \eta$        &  2.1  \\
$\Omega_b^-\to \Xi^{*\,-} \eta'$       &  9.5  \\
$\Omega_b^-\to \Xi^{*\,-} \rho^0$      & 257.3 \\
$\Omega_b^-\to \Xi^{*\,-} \omega$      & 213.9 \\
\hline
\end{tabular}
\label{tab:rates_buud}
\end{center}

\end{table}
\begin{table}

  \begin{center}
\caption{Branching ratios of nonleptonic two-body decays of heavy baryons 
(in units of $10^{-8}$): 
$b \to u;\,\, c \to d$ quark-level transition.} 

\vspace*{.15cm}

\def\arraystretch{1.2}
\begin{tabular}{l|c}
\hline
Mode & Our results \\
\hline
$\Xi_b^0     \to \Sigma^+ D^-$                & 13.6  \\
$\Xi_b^0     \to \Sigma^+ D^{* \,-}$          & 29.6  \\
$\Xi_b^-  \to \Sigma^0 D^-$                   &  7.2  \\
$\Xi_b^-  \to \Sigma^0 D^{*\,-}$              & 15.6  \\
$\Omega_b^-  \to \Xi^0 D^{-}$                 &  1.6  \\
$\Omega_b^-  \to \Xi^0 D^{*\,-}$              &  2.6  \\
$\Omega_b^-  \to \Xi^{*\,0} D^{-}$            &  6.7  \\
$\Omega_b^-  \to \Xi^{*\,0} D^{*\,-}$         & 18.5  \\
$\Omega_b^-  \to \Xi^{-} \bar D^{0}$          & $ 7.3 \times 10^{-2}$ \\
$\Omega_b^-  \to \Xi^{-} \bar D^{*\,0}$       & $13.2 \times 10^{-2}$ \\
$\Omega_b^-  \to \Xi^{*\,-} \bar D^{0}$       &  6.8  \\
$\Omega_b^-  \to \Xi^{*\,-} \bar D^{*\,0}$    & 18.9  \\
\hline
\end{tabular}
\label{tab:rates_bucd}
\end{center}

\begin{center}
\caption{Branching ratios of nonleptonic two-body decays of heavy baryons 
(in units of $10^{-9}$): 
$b \to u; \,\, u \to s$ quark-level transition.} 

\vspace*{.15cm}

\def\arraystretch{1.2}
\begin{tabular}{l|c}
\hline
Mode & Our results \\
\hline
$\Xi_b^-\to \Xi^- \pi^0$         & 0.8 \\    
$\Xi_b^-\to \Xi^- \eta$          & 0.5 \\   
$\Xi_b^-\to \Xi^- \eta'$         & 1.3 \\   
$\Xi_b^-\to \Xi^- \rho^0$        & 2.8 \\
$\Xi_b^-\to \Xi^- \omega$        & 2.1 \\
$\Omega_b^-\to \Omega^- \pi^0$   & 0.7 \\
$\Omega_b^-\to \Omega^- \eta$    & 0.5 \\
$\Omega_b^-\to \Omega^- \eta'$   & 1.1 \\
$\Omega_b^-\to \Omega^- \rho^0$  & 9.1 \\
$\Omega_b^-\to \Omega^- \omega$  & 7.6 \\
\hline
\end{tabular}
\label{tab:rates_buus}
\end{center}

\begin{center}
\caption{Branching ratios of nonleptonic two-body decays of heavy baryons 
(in units of $10^{-2}$): $c \to s;\,\, d \to u$ quark-level transition.} 

\vspace*{.15cm}

\def\arraystretch{1.25}
\begin{tabular}{l|c|c}
\hline
Mode & Our results & Theory  \\
\hline
$\Xi_c^+ \to \Xi^{*\,0} \pi^+$             &   $8.1 \times 10^{-6}$    & \\ 
$\Xi_c^+ \to \Xi^{*\,0} \rho^+$            &   $3.9 \times 10^{-6}$    & \\ 
$\Omega_c^0 \to \Omega^- \pi^+$            &   0.2 & 1.0~\cite{Cheng:1996cs} \\
$\Omega_c^0 \to \Omega^- \rho^+$           &   1.9 & 3.6~\cite{Cheng:1996cs} \\
$\Omega_c^0 \to \Xi^{*\,0}\,\bar K^0$      &   0.3 & \\
$\Omega_c^0 \to \Xi^{*\,0}\,\bar K^{*\,0}$ &   2.0 & \\
\hline
\end{tabular}
\label{tab:rates_csud}
\end{center}

\end{table}

In the future we plan to present predictions also for the nonleptonic modes 
of heavy baryons which are also contributed to by the $W$--exchange diagrams. 
In Ref.~\cite{Ivanov:1997ra} we have shown that for some of the heavy-to-light
transitions 
the total contribution of the $W$--exchange diagrams 
amount up to approximately 60\% of the tree diagram contribution in amplitude, 
and up to approximately 30\% for $b \to c$ transitions.

\subsubsection{The decays 
$\Xi_{b}^{-}(\Xi_{b}^{0})\to\Xi^{-}(\Xi^{0})J/\psi$ and
$\Lambda_{b}\to \Lambda J/\psi$}

The CKM favored decay $\Xi_{b}^{-}\to\Xi^{-}J/\psi$ was first seen by the
CDF~\cite{Aaltonen:2007ap,Aaltonen:2009ny} and the D0 
collaborations~\cite{Abazov:2007am} in the chain of decays 
$\Xi_{b}^{-} \to \Xi^{-} 
J/\psi$ and $J/\psi\to \mu^{+}\mu^{-}$,
$\Xi^{-}\to \Lambda  \pi^{-}$ and $\Lambda \to p \pi^{-}$. These measurements
led to a first determination
of the mass and the lifetime of the $\Xi_{b}^{-}$. PDG 2018 quotes a mass
value of $5794.5\pm1.4$ MeV as a weighted average from the two 
experiments~\cite{PDG:2016}. 
In a later experiment
the LHCb Collaboration collected much higher statistics
on the same decay chain~\cite{Aaij:2014sia}. 
The lifetime of the $\Xi_{b}^{-}$ was determined to be
$\tau(\Xi_{b}^{-})=1.571 \pm 0.040 $ ps~\cite{PDG:2016}. 
This is the most accurate determination of the lifetime of the 
$\Xi_{b}^{-}$ to date. Unfortunately a reliable estimate 
of the branching ratio of the 
decay  $\Xi_{b}^{-} \to \Xi^{-} J/\psi$ is still missing. The decay
$\Xi_{b}^{0}\to\Xi^{0}J/\psi$ is closely related to the decay
$\Xi_{b}^{-} \to \Xi^{-} J/\psi$ by isospin but has not been seen yet. 

It is interesting to note that the neutral current transition 
$\Xi_{b}^{-}\to\Xi^{-}$ can be 
related to the neutral current $\Lambda_{b}\to\Lambda$ 
and charged current $\Lambda_{b}\to p$ transitions
by invoking $SU(3)$ symmetry. This can be 
seen by using the $\bar 3 \otimes 3\,\to 8$ 
Clebsch-Gordan table listed in Ref.~\cite{Kaeding:1995vq}. 
Based on the observation that the antisymmetric 
$[sd]$ and $[ud]$ diquarks are the $(Y=-1/3,I=1/2)$ and 
$(Y=2/3,I=0)$ members of the $\bar 3$ multiplet one needs the
C.G. coefficients~\cite{Kaeding:1995vq}.
\begin{eqnarray}
\label{clebsch}
\mbox{$\Xi_{b}^{-} \to  \Xi^{-}$}&:&
\qquad <\mbox{\boldmath $\bar 3$},-\tfrac 1 3,\tfrac12,-\tfrac12;
\,\mbox{\boldmath $3$} ,
-\tfrac 2 3,0,0|\,\mbox{\boldmath $8$},-1,\tfrac12,-\tfrac12>\,=1\,
\nonumber \\
\mbox{$\Lambda_{b}\to\Lambda$}&:&
\qquad <\mbox{\boldmath $\bar 3$},\tfrac 2 3,0,0;\,\mbox{\boldmath $3$} ,
-\tfrac 2 3,0,0|\,\mbox{\boldmath $8$},0,0,0>\,=\sqrt{2/3}\, , \\
\mbox{$\Lambda_{b}\to p$}&:&
\qquad <\mbox{\boldmath $\bar 3$},\tfrac 2 3,0,0;\, 
\mbox{\boldmath $3$},\tfrac 1 3,
\tfrac 1 2,\tfrac 1 2|\,
\mbox{\boldmath $8$},1,\tfrac 1 2,\tfrac 1 2>\,\,\,=1\,.
\nonumber
\end{eqnarray}
The labeling in~(\ref{clebsch}) proceeds according to the sequence 
$|\,\mbox{\boldmath $R$},Y,I,I_{z}>$
where $\mbox{\boldmath $R$}$ denotes the relevant $SU(3)$ 
representation.
One does not expect that $SU(3)$ breaking effects are large 
at the scale of the bottom baryons.

The angular decay distribution for the cascade decay 
$\Xi_{b}^{-} \to \Xi^{-}(\to \Lambda\,\pi^{-}) \,+\, 
J/\psi(\to \ell^{+}\ell^{-})$
can be adapted from~\cite{Gutsche:2013oea} where now one has to use 
the asymmetry parameter $\alpha_{\Xi}=-0.458\pm0.012$ for the decay
$\Xi^{-}\to \Lambda\,\pi^{-}$~\cite{PDG:2016}. 
As in Ref.~\cite{Gutsche:2013oea} we introduce the following 
combinations of helicity amplitudes 
\bea 
 H_U &=& |H_{\frac{1}{2}1}|^2 +  |H_{-\frac{1}{2}-1}|^2 
\qquad { \rm transverse \,unpolarized}\,, 
\nn
 H_L &=& |H_{\frac{1}{2}0}|^2 +  |H_{-\frac{1}{2}0}|^2
\qquad \, \hfill\mbox{ \rm longitudinal unpolarized}\,, 
\nn 
H_P &=& |H_{\frac{1}{2}1}|^2  -  |H_{-\frac{1}{2}-1}|^2 
\qquad \!\! \! \hfill\mbox{ \rm transverse parity--odd polarized}\,, 
\nn 
H_{L_{P}} &=& |H_{\frac{1}{2}0}|^2 -  |H_{-\frac{1}{2}0}|^2
\qquad \hfill\mbox{ \rm longitudinal polarized}\,, 
\nn
H_S &=& |H_{\frac{1}{2}t}|^2 +  |H_{-\frac{1}{2}t}|^2
\qquad \hfill\mbox{ \rm scalar}\,.
\ena   
Following~\cite{Hrivnac:1994jx,Aaij:2013oxa,Gutsche:2013oea} we also introduce 
linear combinations of normalized squared helicity amplitudes 
$ |\widehat H_{\lambda_{2} \lambda_{V}}|^2$ by writing 
\bea
\label{linear}
\alpha_{b} &=& 
  |\widehat H_{+\tfrac12 0}|^2 - |\widehat H_{-\tfrac12 0}|^2 
+ |\widehat H_{-\tfrac12 -1}|^2 - |\widehat H_{+\tfrac12 +1}|^2 = 
\widehat H_{L_P} - \widehat H_{P}
\,,\nn
r_0 &=& |\widehat H_{+\tfrac12 0}|^2 + |\widehat H_{-\tfrac12 0}|^2 
= \widehat H_{L} \,,\nn 
r_1 &=& |\widehat H_{+\tfrac12 0}|^2 - |\widehat H_{-\tfrac12 0}|^2 
= \widehat H_{L_P}
\,,
\label{eq:asym-param}
\ena
where $ |\widehat H_{\lambda_{2} \lambda_{V}}|^2=
|H_{\lambda_{2} \lambda_{V}}|^2/{\cal H}_N$ 
and where the normalization factor ${\cal H}_N$ is given by 
\bea 
{\cal H}_N \equiv |H_{+\tfrac12 0}|^2 + |H_{-\tfrac12 0}|^2
        + |H_{-\tfrac12 -1}|^2 + |H_{+\tfrac12 +1}|^2\,. 
\ena 

The full joint angular decay distribution of the cascade decay
$B_{1}\to B_{2}(\to B_{3}+P)+V(\to\ell^{+}\ell^{-})$ including the 
polarization of the parent baryon and including lepton
mass effects has been given in~\cite{Gutsche:2013oea}.
The angular decay distribution involves the longitudinal and transverse
polarization components for the daughter baryon $B_{2}$ defined by
\begin{eqnarray}
\label{polzx}
P_{z}(B_{2})&=&|\widehat H_{+\tfrac12 0}|^2 - |\widehat H_{-\tfrac12 0}|^2 
+ |\widehat H_{+\tfrac12 +1}|^2 - |\widehat H_{-\tfrac12 -1}|^2 
= \widehat H_{L_P} + \widehat H_{P} \;,
\nonumber \\
P_{x}(B_{2})&=&2 \,{\rm Re}\Big(\widehat H_{\tfrac12\, 1}
\widehat H^{\ast}_{-\tfrac12 \,0}
+\widehat H_{\tfrac12\, 0}\widehat H^{\ast}_{-\tfrac12\, -1}\Big) \,. 
\end{eqnarray}

Note that one can express the longitudinal polarization of the daughter 
baryon in terms of the linear combinations $r_{1}$ and $\alpha_{b}$ 
in~(\ref{linear}), i.e. one has 
$P_z(B_{2})=2r_{1}-\alpha_{b}=\widehat H_{L_P} + \widehat H_{P}$. The 
magnitude of the transverse polarization $P_{x}(B_{2})$ determines the
size of the azimuthal correlations between the baryon-side and lepton-side
decay plains.

Instead of listing the full angular decay distribution including azimuthal 
terms we shall only list the two-fold polar
angle decay distribution given by
\begin{eqnarray}
\label{angdist1}
W(\theta_3,\theta_\ell)&=&\frac 14\,\Big(1+P_{z}(B_{2})\alpha_{B_{2}}
\cos\theta_3\Big) \mbox{\em v}\cdot(1+2\varepsilon)
+\frac 14\Big((1-3 r_0)(3\cos^2\theta_{\ell}-1) 
\nonumber \\     
&&      - (\alpha_b + r_1)  \alpha_{B_{2}}    
      (3\cos^2\theta_{\ell}-1)\cos\theta_3\Big)     
      \mbox{\em v}\,\cdot\mbox{\em v}^{2}
\end{eqnarray}
$\alpha_{B_{2}}$ denotes the asymmetry factor in the decay 
$B_{2}\to B_{3}+P$, a quantity known from experiment in many cases. One can 
reexpress the two angular parameters 
$(1-3r_{0})$ and $(\alpha_{b}+r_{1})$ in~(\ref{angdist1}) through the
components $\widehat H_{I}$ by writing
$(1-3r_{0})=(\widehat H_{U}-2\widehat H_{L})$ and 
$(\alpha_{b}+r_{1})=(2\widehat H_{L_{P}}-\widehat H_{P})$.
Note that we have kept the overall lepton velocity factor
$v=\sqrt{1-4m^{2}_{\ell}/q^{2}}$ in~(\ref{angdist1}) which results from the 
momentum factor in the decay formula for $V \to \ell^{+}\ell^{-}$. The
trigonometric function $(3\cos^2\theta_{\ell}-1)$ appearing in
Eq.~(\ref{angdist1}) integrates to zero upon $\cos^2\theta_{\ell}$ integration.
In the literature one frequently finds
$(3\cos^2\theta_{\ell}-1)=2\,P_2(\cos\theta_{\ell})$ written in
terms of the Legendre polynomial $P_2(\cos\theta_{\ell})$ of second degree.

We define partial helicity widths in terms of the following bilinear
helicity expressions  
\be 
\Gamma_I = \frac{G_F^2}{32 \pi} \, \frac{|{\bf p}_2|}{M_1^2} \, 
|V_{ij} V^\dagger_{kl}|^2 \, C_{\rm eff}^2 \, f_V^2 \, M_V^2 \, H_I\,, 
\quad\quad I = U, L, P, L_P \,.  
\en 
The total $B_{1} \to B_{2}\,V$ decay width is given by 
\be
\Gamma(B_{1} \to B_{2}V) = \Gamma_U + \Gamma_L \,. 
\en 

We shall also be interested in the cascade decays
$B_{1}\to B_{2}(\to B_{3}\,P)+V(\to P\,P)$ as, e.g., in
$\Lambda_{b}\to \Lambda_{c}(\to p\,                                           
  \bar K^{0})+ D_{s}^{*}(\to D\,\pi)$. The two-fold polar angle
distribution can be calculated to be
\bea
4W(\theta_{3},\theta_{P})=(1+P_{z}(B_{2})\alpha_{B_{2}}
\cos\theta_{3}+(\widehat H_{U}+\tfrac 12 \widehat H_{L})
(3\cos^{2}\theta_{P}-1)+(\widehat H_{L_{P}}+\tfrac 12\widehat H_{P}) 
\alpha_{B_{2}}\cos\theta_{3}(3\cos^{2}\theta_{P}-1)
\ena
Moduli squared of normalized helicity amplitudes 
and asymmetry parameters for the cascade decay 
$\Lambda_{b}\to \Lambda_{c}(\to p\,\bar K^{0})+ D_{s}^{*}(\to D\,\pi)$ 
are shown in Table~\ref{tab:Lambda_cascade_helicity_list}. 

\begin{table}[htbp]

\caption{Moduli squared of normalized helicity amplitudes 
and asymmetry parameters for the cascade decay 
$\Lambda_{b}\to \Lambda_{c}(\to p\,\bar K^{0})
+ D_{s}^{*}(\to D\,\pi)$ transition.}
\begin{tabular}{r|l||r|l||r|l}
\hline
Quantity & Our results &
Quantity & Our results &
Quantity & Our results \\

\hline
$\tH{+\tfrac12 +1}$ & 4.46 $\times 10^{-2}$ 
& $\widehat H_U$      & 0.358 
& $\alpha_b$ & -0.364 \\    
\hline 
$\tH{+\tfrac12  0}$ & 4.98 $\times 10^{-3}$ 
& $\widehat H_L$      & 0.642 
& $r_0$      &  0.642 \\
\hline 
$\tH{-\tfrac12  0}$ & 0.637 
& $\widehat H_P$      & $-0.268$ 
& $r_1$      & -0.632 \\
\hline 
$\tH{-\tfrac12 -1}$ & 0.313 
& $\widehat H_{L_{P}}$& $-0.633$ 
& $(P_x, P_z)$ & (-0.901, 0.416) \\
\hline
\end{tabular}
\label{tab:Lambda_cascade_helicity_list}
\end{table}

Next we turn to the decay $\Lambda_{b}\to \Lambda J/\psi$. The angular 
decay distribution of the cascade chain 
$\Lambda_{b}\to \Lambda(\to p\,\pi^{-}) \,+\, J/\psi(\to \ell^{+}\ell^{-})$
including polarization effects of the parent baryon $\Lambda_{b}$ 
has been discussed in detail in~\cite{Bialas:1992ny,%
Hrivnac:1994jx,Gutsche:2013pp}. 
A new insight has been gained concerning the accessibility of the 
polarization of the produced parent baryon $\Lambda_{b}$~\cite{Aad:2014iba}.
In a symmetric collider such as the $pp$-collider where the sample of
$\Lambda_{b}$'s are produced over a symmetric interval of pseudorapidity the
polarization of the $\Lambda_{b}$ must average to zero. In order to
access the polarization of the $\Lambda_{b}$ one must e.g. divide the
pseudorapidity interval into two forward and backward hemispheres which has
not been done in the LHCb~\cite{Aaij:2013oxa} and ATLAS~\cite{Aad:2014iba} 
experiments. We shall therefore
no longer discuss the polarization dependent terms in the angular decay
distribution. Such a measurement has to wait for a future analysis.

\begin{table}[htbp]

\caption{Moduli squared of normalized helicity amplitudes 
and asymmetry parameters for the 
$\Lambda_{b}\to \Lambda(\to p\,\pi^{-}) J/\psi(\to \ell^{+}\ell^{-})$ 
transition.}
\begin{tabular}{r|l|r|r|r}
\hline
Quantity & Our results & 
LHCb~\cite{Aaij:2013oxa} \quad\quad & 
ATLAS~\cite{Aad:2014iba} \quad\quad & 
CMS~\cite{CMS:2016iaf}   \quad\quad \\
\hline
$\tH{+\tfrac12 +1}$ & 2.34 $\times 10^{-3}$ & -0.10 $\pm$ 0.04 $\pm$ 0.03 & 
$(0.08^{+0.13}_{-0.08} \pm 0.06)^2$ &  0.05 $\pm$ 0.04 $\pm$ 0.02 \\
$\tH{+\tfrac12  0}$ & 3.24 $\times 10^{-4}$ &  0.01 $\pm$ 0.04 $\pm$ 0.03 
& $(0.17^{+0.12}_{-0.17} \pm 0.09)^2$ & -0.02 $\pm$ 0.03 $\pm$ 0.02 \\
$\tH{-\tfrac12  0}$ & 0.532              &  0.57 $\pm$ 0.06 $\pm$ 0.03 
& $(0.59^{+0.06}_{-0.07} \pm 0.03)^2$ &  0.51 $\pm$ 0.03 $\pm$ 0.02 \\
$\tH{-\tfrac12 -1}$ & 0.465              &  0.51 $\pm$ 0.05 $\pm$ 0.02 
& $(0.79^{+0.04}_{-0.05} \pm 0.02)^2$ &  0.46 $\pm$ 0.02 $\pm$ 0.02 \\[1ex]
\hline
$\alpha_b$ & -0.069 &  0.05 $\pm$ 0.17 $\pm$ 0.07 & 
0.30 $\pm$ 0.16 $\pm$ 0.06 & -0.12 $\pm$ 0.13 $\pm$ 0.06 \\
$r_0$      &  0.533 &  0.58 $\pm$ 0.02 $\pm$ 0.01 &                        & \\
$r_1$      & -0.532 & -0.56 $\pm$ 0.10 $\pm$ 0.05 &                        & \\
$P_z(\Lambda)$ & -0.995 & -1.17 $\pm$ 0.26 $\pm$ 0.12 & & \\
$P_x(\Lambda)$ &  0.095 & & & \\
\hline
\end{tabular}
\label{tab:Lambda_helicity_list}
\end{table}

We list some numerical results of our previous analysis on the decay 
$\Lambda_{b}\to \Lambda J/\psi$ in Table~\ref{tab:Lambda_helicity_list}. 
Table~\ref{tab:Lambda_helicity_list} shows that the longitudinal 
polarization of the daughter baryon
$\Lambda$ is almost $100\,\%$ and negative. In fact, one finds 
$P_{z}(\Lambda)=\widehat H_{L_P} + \widehat H_{P}=-0.995$. This leaves very 
little room for the transverse polarization $P_{x}(\Lambda)$ since 
$P^{2}_{z}(\Lambda)+P^{2}_{x}(\Lambda)=1$. We set the $T$-odd 
polarization component $P_{x}(\Lambda)$ to zero since the helicity amplitudes
are nearly real in our analysis. From the numbers in 
Table~\ref{tab:Lambda_helicity_list} 
and using the definition~(\ref{polzx}) one finds $P_{x}(\Lambda)=0.095$. 
A small transverse polarization $P_{x}(\Lambda)$ in turn implies small 
azimuthal correlations between the two plains defined by $(p,\,\pi^{-})$ and
$ (\ell^{+},\ell^{-})$ as was found in~\cite{Gutsche:2013oea} and which has 
been confirmed by the azimuthal measurement in~\cite{Aad:2014iba}.  

For the asymmetry parameter $\alpha_{b}$ we obtained the small 
negative value $\alpha_{b}\simeq -0.07$~\cite{Gutsche:2013pp}. 
The value of $\alpha_{b}$ determines the analyzing power of the cascade 
decay $\Lambda_{b}\to \Lambda(\to p\,\pi^{-}) \,+\, J/\psi(\to \ell^{+}\ell^{-})$
to measure the polarization of the parent baryon $\Lambda_{b}$. A small
value of $\alpha_{b}$ implies a poor analyzing power of such an experiment.
Our calculated value for the asymmetry parameter $\alpha_{b}=-0.07$
agrees within one standard deviation with the slightly revised
value of the asymmetry parameter $\alpha_{b}=0.05\pm0.17\pm0.07$ reported 
in~\cite{Aaij:2013oxa}. A new measurement on $\alpha_{b}$ was reported by the 
ATLAS collaboration~\cite{Aad:2014iba}. Our result is more than two
standard deviations away from their reported value of  
$\alpha_{b}=0.30\pm0.16\pm0.06$. We mention that most of the theoretical
model calculations predict small negative values for $\alpha_{b}$ ranging
from (-0.09) to (-0.21) (see
Table VI in~\cite{Gutsche:2013oea}) except the calculation 
of~\cite{Leitner:2004dt,Ajaltouni:2004zu}
who quote a value of $\alpha_{b}=0.777$. We believe the calculation 
of~\cite{Leitner:2004dt,Ajaltouni:2004zu} to be erroneous. The relation
between the helicity amplitudes and the HQET form factors $F_{1}$ and 
$F_{2}$ given in~\cite{Leitner:2004dt,Ajaltouni:2004zu} is not correct. 
They also obtain a polarization of the daughter baryon $\Lambda$ 
of $P_{z}(\Lambda)=-9\%$ which is much smaller than what is
obtained in the other model calculations.

\begin{figure}
\begin{center}
\epsfig{figure=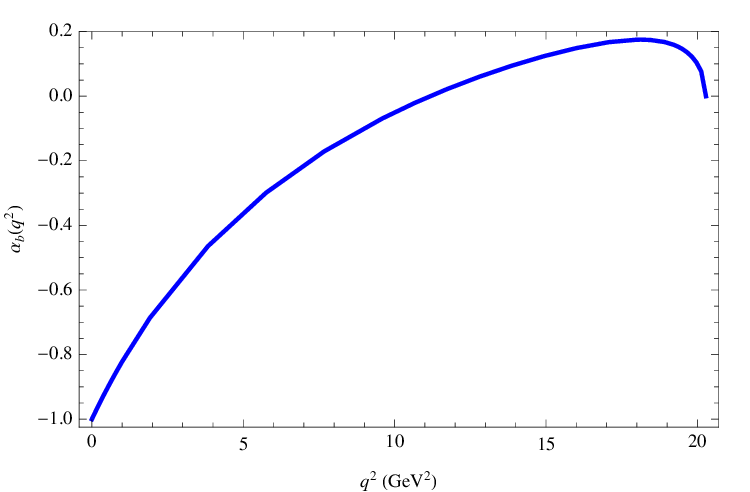,scale=.65}
\end{center}
\noindent
\caption{$q^2$ dependence of the asymmetry parameter $\alpha_b(q^2)$ 
in the $\Lambda_{b}\to \Lambda$ transition. 
\label{fig:alphaB}}
\end{figure}

In Fig.~\ref{fig:alphaB} we show a plot of $\alpha_{b}$ 
over the whole range of accessible
$q^{2}$-values. At the kinematical limits $q^{2}=0$ and 
$q^{2}=(M_1-M_2)^2=20.28$ GeV$^2$ $\alpha_b$ takes the values
$\alpha_{b}=-1$ and $\alpha_{b}=0$ in agreement with the corresponding 
limiting values of Eq.~(\ref{eq:asym-param}). 
The asymmetry parameter $\alpha_{b}$ 
goes through zero between $q^{2}=m^{2}_{J/\psi} = 9.59$ GeV$^2$ 
and $q^{2}=m^{2}_{\psi(2S)} = 13.59$ GeV$^2$ 
as demonstrated in the numerical results of~\cite{Gutsche:2013pp}. We show
this plot in order to demonstrate that it is very unlikely that the
asymmetry value reaches a large positive value as 
in~\cite{Leitner:2004dt,Ajaltouni:2004zu} between the above two 
model independent limiting values.    

\subsubsection{The decays $\Lambda_{b}^{0}\to \Lambda_{c}^{+} D_{s}^{-}$ and
$\Lambda_{b}^{0}\to \Lambda_{c}^{+} D^{-}$}

Both the Cabibbo-favored decay 
$\Lambda_{b}^{0}\to \Lambda_{c}^{+} D_{s}^{-}$ and the Cabibbo suppressed decay
$\Lambda_{b}^{0}\to \Lambda_{c}^{+} D^{-}$ have been seen 
by the LHCb collaboration~\cite{Aaij:2014pha}. The PDG 2018 quotes 
branching fractions of 
${\cal B}(\Lambda_{b}^{0}\to \Lambda_{c}^{+} D_{s}^{-})
=(1.1\pm0.1)\times 10^{-2}$ and 
${\cal B}(\Lambda_{b}^{0}\to \Lambda_{c}^{+} D^{-})
=(4.6\pm0.6)\times 10^{-4}$ for these decays~\cite{PDG:2016}. 

We employ a notation where we label the scalar time component of the
current by the suffix $''t''$ where $\lambda_{t}=0$ in order to
distinguish between the $(J=0)$ and $(J=1)$ transformation of the two 
helicity zero cases.
We shall again define normalized helicity amplitudes by writing
$|\widehat H_{\lambda_{2}\,t}|^{2}=|H_{\lambda_{2}\,t}|^{2}/{\cal H}_N $ where
${\cal H}_N$ is now given by
${\cal H}_N=|H_{\frac{1}{2}t}|^{2}+|H_{-\frac{1}{2}t}|^{2}$.
In this notation the longitudinal polarization of the daughter baryon 
$B_{2}$ is given by
\bea 
P_z(B_{2})= |\widehat H_{\frac{1}{2}t}|^{2}
-|\widehat H_{-\frac{1}{2}t}|^{2} = \left\{
\begin{array}{cc}
-0.989, & \ \Lambda_{b} \to \Lambda_{c} D \ {\rm mode} \\
-0.986, & \ \Lambda_{b} \to \Lambda_{c} D_s \ {\rm mode} \\
\end{array}
\right.
\ena 
One then has the polar decay distribution
\begin{equation}
\label{angdistP}
W(\theta_{3})\propto \left(1+P_z(B_{2})\lambda_{B_{2}}\cos\theta_{3}\right) \;.
\end{equation}
The transverse polarization component $P_x(B_{2})$ does not come into play
in these decays.

\subsubsection{ The $1/2^+ \to 3/2^+ + 1^-$ factorizing decay 
$\Omega_{b}^{-}\to\Omega^{-}J/\psi$}

The $\Omega^{-}_{b}$ has a clear signature through the decay  
$\Omega_{b}^{-}\to\Omega^{-}J/\psi$. This decay mode was first observed 
in~\cite{Aaltonen:2009ny,Abazov:2008qm}. The PDG 2018 quotes 
the average values for the mass $6046.1\pm 1.7$ MeV and the life time 
$\tau(\Omega_{b}^{-})=1.65^{+0.18}_{-0.16}$ ps~\cite{Aaij:2014sia}.  
We shall take the LHCb result for our evaluation of branching rates.

We begin with the decay $1/2^{+} \to 3/2^{+}\,1^-$ where we concentrate on
the decay chains of the daughter baryon $3/2^{+}\to 1/2^{+}+P$ and the
vector meson $V \to \ell^{+}\ell^{-}$.

\begin{table}[!htbp]

\caption{Moduli squared of normalized helicity amplitudes 
and asymmetry parameters for the cascade decay
$\Omega_b\to \Omega^-(\to \Xi^-\,p\,, \Lambda K^-)
+ J/\psi (\to \ell^+\ell^-)$ transitions.}
\begin{tabular}{l|c}
\hline
Quantity & Our results \\
\hline
${\cal B}(\Omega_{b}\to \Omega^- J/\psi)$ 
& 0.08 \% \\
$\widehat {\cal H}_{U_{1/2}}$      & 0.236 \\     
$\widehat {\cal H}_{U_{3/2}}$      & 0.312 \\     
$\widehat {\cal H}_L$              & 0.452 \\
\hline
\end{tabular}
\label{tab:Omega_helicity_list}
\end{table}

In Table~\ref{tab:Omega_helicity_list} 
we list our results on the branching ratio of this decay
together with our values for $\widehat {\cal H}_{U_{1/2}},
\,\widehat {\cal H}_{U_{3/2}}$
and $\widehat {\cal H}_{L}$. Latter values determine the polar angle
distribution of the
subsequent decay $\Omega^{-}\to \Xi\pi,\,\Lambda\,K^{-}$. We have defined
$\widehat {\cal H}_{U_{1/2}}=|\hat H_{\frac 12\,1}|^2+|\hat H_{-\frac 12\,-1}|^2$,
$\widehat {\cal H}_{U_{3/2}}=|\hat H_{\frac 32\,1}|^2+|\hat H_{-\frac 32\,-1}|^2$
and
$\widehat {\cal H}_{L}=|\hat H_{\frac 12\,0}|^2+|\hat H_{-\frac 12\,}|^2$,
where, as before, $|\hat H_{\lambda_2\,\lambda_V}|^2=|H_{\lambda_2\,\lambda_V}|^2
/{\cal H}_V $ and ${\cal H}_V= \sum_{\lambda_2\,,\lambda_V}|H_{\lambda_2\,\lambda_V}|^2$.
A much smaller value of the branching fraction has been calculated in
\cite{Fayyazuddin:2017sxq} where the authors predict
${\cal B}(\Omega_{b}\to \Omega^- J/\psi)= 4.5\times 10^{-5}$. The smallness of the
branching ratio in the calculation~\cite{Fayyazuddin:2017sxq} can be traced
to the erroneous assumption that the transition form factors in the process
fall from $q^2=0$ to $q^2=m^2_{J/\psi}$.

We define partial helicity widths in terms of the following bilinear
helicity expressions  
\be 
\Gamma_I = \frac{G_F^2}{32 \pi} \, \frac{|{\bf p}_2|}{M_1^2} \, 
|V_{ij} V^*_{kl}|^2 \, C_{\rm eff}^2 \, f_V^2 \, M_V^2 \, {\cal H}_I\,, 
\quad\quad I = U_{1/2},\, U_{3/2},\, L \,.  
\en 
The $B_{1} \to B_{2}V$ decay width is given by 
\be
\Gamma(B_{1}(1/2^{+}) \to B_{2}(3/2^{+})\,V) 
= \Gamma_{U_{1/2}} + \Gamma_{U_{3/2}}+ \Gamma_L \,. 
\en

The two-fold joint angular decay distribution for the cascade decay
$1/2^{+}\to 3/2^{+}(\to 1/2^{+}\,0^{-})+V(\to \ell^{+}\ell^{-})$ is
simplified by the observation that the relevant decays of the daughter
baryon  with $B_{2}(3/2^{+})\to B_{3}(1/2^{+})\,0^{-}$ 
and $\Omega^{-}\to \Xi\,\pi, \Lambda\,K^{-}$ 
are almost purely parity conserving~\cite{PDG:2016}, 
i.e. for the decay $3/2^{+}\to1/2^{+}\,0^{-}$ one has 
$H_{\lambda_{3}0}=H_{-\lambda_{3}0}$. This also follows 
from a simple quark model calculation~\cite{koe72,Korner:1992wi}. 
The fact that the vector current induced transition 
$3/2^{+} \to 1/2^{+}$ is conserved
in the simple quark model can be traced to the approximation 
that there is no spin interaction between the two light spectator 
quarks~\cite{Hussain:1990uu}. This means there will be no parity-odd
linear $\cos\theta_{B_3}$ terms in the angular decay distribution.

The two-fold joint polar angle decay distribution can be derived 
from the master formula 
\bea
W(\theta_\ell,\theta_B)\propto  
\sum_{\rm helicities}
|h^{V}_{\lambda_{\ell^{+}}\lambda_{\ell^{-}}}|^2
\left[d^1_{\lambda_V,\lambda_{\ell^{+}}
-\lambda_{\ell^{-}}}(\theta_{\ell})\right]^2
|H_{\lambda_{2}\lambda_{V}}|^2
\left[d^{3/2}_{\lambda_{2}\lambda_3}(\theta_B)\right]^2 , 
\label{master}
\ena
where the summation extends over all possible helicities
$\lambda_{\ell^{+}},\lambda_{\ell^{-}},\lambda_{2},\lambda_3
=\pm \tfrac12$ and $\lambda_V=0,\pm 1$. In~(\ref{master}) we have left
out the overall factor $|H_{\lambda_{3}0}|^{2}$. The required table of the 
Wigner symbol $d^{3/2}_{mm'}(\theta)$ can
be found e.g. in~\cite{Varshalovich:1988ye}.
 
The vector current leptonic helicity amplitudes for the decay 
$V\to \ell^{+}\ell^{-}$ are given by 
(see \cite{Gutsche:2013pp})
\begin{equation}
\mbox{flip:}\quad h^{V}_{-\tfrac12 -\tfrac12} = 
h^{V}_{+\tfrac12 +\tfrac12} =2m_l\,, \qquad \mbox {nonflip:}\quad
h^{V}_{-\tfrac12 +\tfrac12} = h^{V}_{+\tfrac12 -\tfrac12} =\sqrt{ 2 q^2}\,.
\label{flipnoflip}
\end{equation}
For the angular decay distribution one obtains
\begin{eqnarray}
\label{angdist3}
W(\theta_{B_3},\theta_{\ell})&\propto&\tfrac38\,v
\Big((1+\cos^{2}\theta_{\ell})v
+8\varepsilon\Big)\tfrac14\Big(\widehat {\cal H}_{U_{1/2}}
(1+3\cos^{2}\theta_{B_{3}})+\widehat {\cal H}_{U_{3/2}}3\sin^{2}\theta_{B_{3}}\Big)
\nonumber \\
&&+\,\widehat {\cal H}_{L}\tfrac38\,v\Big((1+\cos^2\theta_{\ell})\,v
+8\Big)\tfrac14(1+3\cos^2\theta_{B_3})
\end{eqnarray}
We have again included an overall factor of $v=(1-4\varepsilon)^{1/2}$
from the momentum factor in the decay formula for the decay 
$V \to \ell^{+}\ell^{-}$.

As in Eq.~(\ref{angdist1}) we can represent the angular decay distribution 
in an equivalent form in terms of the second order Legendre polynomials
$P_2(x)=(3x^2-1)/2$ with $x=cos\theta_{B_{3}},\cos^\theta_{\theta_\ell}$.
Using the same normalization as 
in~(\ref{angdist3}) one has
\begin{eqnarray}
\label{angdist4}
4W(\theta_{B_{3}},\theta_{\ell})&\propto&v\,(1+2\varepsilon)
+\tfrac12(3\cos^{2}\theta_{B_{3}}-1)
\Big(\widehat {\cal H}_{U_{1/2}}-\widehat {\cal H}_{U_{3/2}}+\widehat {\cal H}_{L}\Big)
v(1+2\varepsilon)+\tfrac{1}{4}(3\cos^{2}\theta_{\ell}-1)v^{3}
\nonumber \\
&&+ \, \tfrac{1}{8}(3\cos^{2}\theta_{B_{3}}-1)(3\cos^{2}\theta_{\ell}-1)
\Big( {\cal H}_{U_{1/2}}
-\widehat {\cal H}_{U_{3/2}}+\widehat {\cal H}_{L}\Big)v^{3}
\end{eqnarray}
such that the angle dependent parts integrates to zero
\cite{Hrivnac:1994jx}.
Upon integration one has
\begin{eqnarray} 
\int d\cos\theta_{\ell}d\cos\theta_{B}W(\theta_{B},\theta_{\ell})
=\Big(\widehat {\cal H}_{U_{1/2}}+\widehat {\cal H}_{U_{3/2}}
+\widehat {\cal H}_{L}\Big) \  v\,(1+2\varepsilon\big)=v\,(1+2\varepsilon\big) \;.
\end{eqnarray}

Next we discuss the decays $1/2^{+}\to 3/2^{+}\,0^{-}$. Similar to the
current induced transition $3/2^{+}\to 1/2^{+}$ one finds from a naive quark
model calculation~\cite{Korner:1992wi} 
that the vector current transition $<3/2^{+}|J_{\mu}^{V}|1/2^{+}>$ 
is conserved. The basic mechanism is the same as mentioned in the discussion
of the $\langle 1/2^{+}|J_{\mu}^{V}|3/2^{+}\rangle$ transition. The crucial
assumption is that there is no spin interaction between the spectator
quarks. The conservation of the vector current 
implies that the helicity amplitudes $H^V_{\pm\lambda_2,t}$ vanish, i.e.
one has  $H^V_{\pm\lambda_2,t}=0$. We also find vector current conservation
$q^\mu \langle 3/2^{+}|J_{\mu}^{V}|1/2^{+}\rangle=0$ in our more sophisticated
CCQM quark model, i.e. one has again $H^V_{\pm\lambda_2,t}=0$. The vector current
conservation in the CCQM can be traced to the structure of the interpolating
currents of the $\Omega_c$ and $\Omega^-$ of Table I.

The rate can be
calculated from
\bea 
\Gamma(1/2^{+}\to 3/2^{+}\,0^{-})= 
\frac{G_F^2}{32 \pi} \, \frac{|{\bf p}_2|}{M_1^2} \, 
|V_{ij} V^\dagger_{kl}|^2 \, C_{\rm eff}^2 \, f_P^2 \, M_P^2 \, {\cal H}_S\,, 
\ena 
where ${\cal H}_S= |H_{\frac 12\,t}|^2+|H_{-\frac 12\,t}|^2=2\,|H_{\frac 12\,t}|^2$.
The corresponding
branching ratio for $\Omega^-_b \to \Omega^-\,\eta_c$ is predicted to be
${\cal B}=5.0 \times 10^{-4}$ (see Table III).

The baryon-side decay distribution
in this decay is identical to that of the longitudinal contribution in
Eq.~(\ref{angdist3}), i.e. one has
\begin{equation}
\label{angdist44}
W(\theta_{B_{3}})=\tfrac 14(1+3\cos\theta_{B_3})=  \tfrac 12(1+\tfrac 12
(3\cos\theta_{B_3}-1))=\tfrac 12 (1+P_2(\cos\theta_{B_3}))\,.
\end{equation}

\subsubsection{The decay $\Lambda_c \to p\,+\,\phi$}

The Cabibbo-suppressed charm baryon decay  
$\Lambda_c \to  p \phi$ was first observed by the 
ACCMOR Collaboration in the NA32 experiment at CERN~\cite{Barlag:1990yv} 
with a branching ratio of 
$B(\Lambda_c \to p \phi)/B(\Lambda_c \to p K^- \pi^+) = 0.04 \pm 0.03$. 
Three years later the E687 Collaboration 
Collaboration at Fermilab reported an 
upper limit for its branching relative to the mode 
$B(\Lambda_c \to p \phi)/B(\Lambda_c \to p K^- K^+) < 0.58$ 
at 90\% C.L.~\cite{Frabetti:1993ew}. 
In 1996 the CLEO Collaboration~\cite{Alexander:1995hd} confirmed 
the previously observed $\Lambda_c \to  p \phi$ decay mode with significant 
statistics and measured the branching ratios 
$B(\Lambda_c \to p \phi)/B(\Lambda_c \to p K^- \pi^+) 
= 0.024 \pm 0.006 \pm 0.003$ and 
$B(\Lambda_c \to p \phi)/B(\Lambda_c \to p K^- K^+) = 0.62 \pm 0.20 \pm 0.12$. 
Finally, in 2002 the Belle Collaboration at KEK~\cite{Abe:2001mb} 
measured the ratio of branching rates with much higher accuracy 
$B(\Lambda_c \to p \phi)/B(\Lambda_c \to p K^- \pi^+) 
= 0.015 \pm 0.002 \pm 0.002$.

The good news is that the long-standing problem of the absolute branching 
ratio of the decay $\Lambda_c \to p K^- \pi^+$ has been solved by the 
BELLE~\cite{Zupanc:2013iki} and the BES III collaborations
\cite{Ablikim:2015flg}.
Thus the above branching rate ratios can be converted into absolute
branching ratios. 

In the present case the information on the polarization of the daughter
baryon, the proton, is of no particular interest since the polarization
of the proton cannot be probed.
The angular decay 
distribution on the lepton side can be read off from Table I in  
Ref.~\cite{Gutsche:2013oea} or can be obtained by integrating 
Eq.~(\ref{angdist1}) over $\cos\theta_{3}$.
One has ($\varepsilon=m_\ell^2/m^2_\phi$) 
\begin{equation}
\label{angdist2}
W(\theta_{\ell})\propto\,v(1+2\varepsilon)\Big(1+\frac14
\frac{v^{2}}{1+2\varepsilon}(\widehat H_{U}
-2\widehat H_{L})(3\cos^{2}\theta_{\ell}-1)\Big)
\end{equation}
where $\theta_{\ell}$ is the polar angle of either of the leptons 
with respect to 
the momentum direction of the $\phi$ in the $\Lambda_{c}$ rest frame. Note
that the second $\cos\theta_{\ell}$-dependent term vanishes after integration.
The muon mass corrections to the rate term in~(\ref{angdist2}) are quite 
small as is evident
from the expansion $v(1+2\varepsilon)=1-4\varepsilon^{2}+O(\varepsilon^{3})
\approx 0.998$.
The muon mass corrections to the $\cos\theta_{\ell}$ dependent term, however,
are larger and can be seen to amount to
$v^2/(1+2\varepsilon)=1-6\varepsilon + O(\varepsilon^{3}) \simeq 6\,\%$.  

We present our results in Tables~\ref{tab:rates} 
and~\ref{tab:Lcpleptons_rates}
and compare them with data presented by the Particle Data 
Group~\cite{PDG:2016} and the predictions of other theoretical 
approaches~\cite{Ivanov:1997ra,Korner:1992wi,%
Cheng:1991sn,Zenczykowski:1993hw,Datta:1995mn}.  
We calculate the width of the cascade decay 
$\Lambda_c\to p\,+\,\phi(\to \ell^+\ell^-)$
by using the zero width approximation
\be
{\rm B}(\Lambda_c \to p\,+\,\phi(\to \ell^+ \ell^-) ) 
=  
{\rm B}(\Lambda_c \to p\,+\,\phi) \  
{\rm B}(\phi \to \ell^+ \ell^-) \,. 
\label{eq:ZWA}
\en 
We take the value of the leptonic decay constant
$f_{\phi} = 226.6$ MeV from the $e^{+}e^{-}$ mode
calculated in our approach in agreement with data. 
Using the result Eq.~(\ref{angdist2}) we can write down the general
$m_{\ell}\neq 0$ decay formula
\be 
\Gamma(\phi \to \ell^+  \ell^-) = 
\frac{4 \pi \alpha^2}{27} \, \frac{f_{\phi}^2}{m_{\phi}} \, 
\sqrt{1 - \frac{4 m_\ell^2}{m_{\phi}^2}}  
\, \biggl(1 + \frac{2 m_\ell^2}{m_{\phi}^2}\biggr)   
\label{eq:leptonic}
\en 
to evaluate the muonic mode. As already remarked above, muon mass effects are 
negligible in the rate as Table~\ref{tab:Lcpleptons_rates} shows. 
Note that we give more stringent upper limits on the branching ratios of
these decays compared to the data~\cite{PDG:2016}. The numerical value for
the asymmetry $(\hat H_U- 2\hat H_L)$ is 
\be 
\hat H_U- 2\hat H_L = -0.735 \,. 
\en 

\begin{table}[hb]
\begin{center} 
\caption{Branching ratio $B(\Lambda_c \to p \phi)$ in units $10^{-4}$.} 
\vspace*{.15cm}

\def\arraystretch{1.5}
    \begin{tabular}{|c|c|c|}
      \hline
Our result  & Theoretical predictions & Data~\cite{PDG:2016} \\
\hline
14.0 & 19.5~\cite{Cheng:1991sn}; 
      21.5~\cite{Korner:1992wi}; 
      9.89~\cite{Zenczykowski:1993hw}; 
      4.0~\cite{Datta:1995mn}; 
      27.3 $\pm$ 17.9~\cite{Ivanov:1997ra}  
    & $10.8 \pm 1.4$ \\
\hline
\end{tabular}
\label{tab:rates}
\end{center}

  \begin{center}
    \caption{Branching ratios 
$B(\Lambda_c \to p \phi\,(\ell^+\ell^-))$ in units
      of $10^{-7}$.}

    \vspace*{.15cm}
    \def\arraystretch{1.5}
    \begin{tabular}{|c|c|c|}
      \hline
Mode & Our results & Data~\cite{PDG:2016} \\ 
\hline
$\Lambda_c \to p\,+\,e^+ e^-$         &  4.11 & $<$  55 \\
\hline
$\Lambda_c \to p\,+\, \mu^+ \mu^-$    &  4.11 & $<$ 440 \\
\hline
\end{tabular}
\label{tab:Lcpleptons_rates}
\end{center}
\end{table}

\section{Summary} 

We have calculated the decay properties of a number of nonleptonic
heavy baryon decays. Of the many possible nonleptonic
heavy baryon decays we have singled out those decays which proceed solely 
through the tree diagram contribution without a possible contamination from
$W$-exchange contributions. 
Note that we have not included penguin-type operators in our set of operators 
as, e.g., the penguin operator governing the quark-level transition 
$b \to s \bar s s$ 
that induces the decay $\Lambda_b \to \Lambda + \phi$ recently seen by 
the LHCb Collaboration~\cite{Aaij:2016zhm}. 
We plan to study decays induced by penguin-type operators in the future.
For the decays dominated by tree diagram contribution we have calculated
decay rates, branching ratios and asymmetry parameters associated with
the polar angle distributions of their subsequent decay chains.
Some of these decays have been seen which has lead to measurements of their
lifetimes and their masses. We have also provided results for decays with
prominent branching ratios and signatures which
have not been observed up to now. Measuring lifetimes and 
masses is only the first step towards a comprehensive analysis of heavy
baryon decays which would aim for the determination of branching ratios and 
asymmetry parameters in angular
decay distributions as has e.g. been done in the decay 
$\Lambda_{b} \to \Lambda \,J/\psi$. We estimate our errors on branching
fractions to be $\sim 20\,\%$. Our errors on the asymmetry parameters are
much smaller since the errors cancel out in the helicity rate ratios 
that describe the asymmetry parameters.

It would be interesting to extend our analysis to all nonleptonic two-body 
decays of the heavy baryons. Such a calculation would involve also
contributions from the $W$--exchange graphs called
color commensurate (C), exchange (E) and bow-tie (B) graphs 
in~\cite{Leibovich:2003tw}. For the decays of the bottom baryon states the
authors of~\cite{Leibovich:2003tw} derived a hierarchy of strength
$T\gg C,E \gg B$ from a qualitative SCET analysis. It would be interesting to
find out whether the dominance of the tree graph contribution
in the $b \to c$ and
$b \to u$ sector predicted in~\cite{Leibovich:2003tw} is borne out by the 
phenomenology of these decays. In particular, it would be interesting to
determine the size of the $W$--exchange contributions to the decays
$\Lambda_{b}\to \Lambda \eta(\eta')$. 
Within the general class of nonleptonic two-body baryon decays one also finds
decays which proceed solely by $W$-exchange diagrams. Among these are the 
decays in which the transition 
between the two spectator quarks is forbidden due to the fact that one has
a transition between a symmetric and antisymmetric spin-flavor configuration
such as in $[ud] \to \{ud\}$. A few examples are the 
$c\to s\bar d u$ decays $\Lambda^{+}_{c}\to\Delta^{++}K^{-}$,
$\Lambda^{+}_{c}\to\Sigma^{0}\pi^{+}$ and $\Xi_{c}^{0}\to \Omega^{-}K^{+}$.
Then there is a multitude of decays where the flavor composition of the
parent and daughter baryon preclude a tree graph contribution. In the
$(c \to s;\,d \to u)  $ sector two sample decays are
$\Lambda_c^+ \to \Xi^0\,K^+$ and $\Lambda_c^+ \to \Xi^{*0}(1530)\,K^+$ which
have been
observed with the sizable branching ratios of
${\cal B}=(5.90\pm0.86\pm0.39)\times 10^{-3}$
and ${\cal B}=(5.02\pm0.99\pm0.31)\times 10^{-3}$,
respectively~\cite{Ablikim:2018bir}. 
Among the CKM favored $b \to c \bar u d $ decays are
$\Lambda^{0}_{b} \to \Sigma_{c}^{0}\pi^{0}$ and $\Xi_b^- \to \Sigma_c^0 K^-$ as
well as the $b \to c \bar c s $ decays $\Lambda_b^0 \to \Xi_c^+ D^0$, the
CKM suppressed $b \to u \bar u  d$ decays 
$\Xi_{b}^{-} \to \Lambda^{0}\pi^{-}$
and $\Lambda^{0}_{b} \to \Lambda^{0}K^{0}$ 
and the doubly Cabibbo suppressed $b \to u \bar u s$ decays 
$\Xi_{b}^{-} \to \Xi^{0}\pi^{-}$ and $\Lambda^{0}_{b} \to \Xi^{0}K^{0}$.
The latter bottom baryon decays are predicted to be severely suppressed
according to the SCET analysis of~\cite{Leibovich:2003tw}. Again, it 
would be important to find out whether this prediction shows up in the
experimental rates.

A last remark concerns the experimental observability of the decays discussed 
in this paper. The best way to check on the observability of the decays
would be to generate MC events according to the
described decay chains and to process them through the detector. 

\begin{acknowledgments}

This work was funded
by the German Bundesministerium f\"ur Bildung und Forschung (BMBF)
under Project 05P2015 - ALICE at High Rate (BMBF-FSP 202):
``Jet- and fragmentation processes at ALICE and the parton structure          
of nuclei and structure of heavy hadrons'',
Carl Zeiss Foundation under Project ``Kepler Center f\"ur Astro- und 
Teilchenphysik: Hochsensitive Nachweistechnik zur Erforschung des 
unsichtbaren Universums (Gz: 0653-2.8/581/2)'', 
by CONICYT (Chile) PIA/Basal FB0821, by the Russian Federation
program ``Nauka'' (Contract No. 0.1764.GZB.2017), by Tomsk State University
competitiveness improvement program under grant No. 8.1.07.2018, and 
by Tomsk Polytechnic University
Competitiveness Enhancement Program (Grant No. VIU-FTI-72/2017).
M.A.I.\ acknowledges the support from  PRISMA cluster of excellence
(Mainz Uni.). M.A.I. and J.G.K. thank the Heisenberg-Landau Grant for
the partial support.

\end{acknowledgments}

\end{document}